\documentclass[iop, revtex4]{emulateapj}
\usepackage{epstopdf}
\usepackage{multirow}
\usepackage{threeparttable}
\usepackage{natbib}
\usepackage{float}
\newcommand{\unit}[1]{\ensuremath{\, \mathrm{#1}}}

\newcommand{\Msol}{\hbox{$\mathrm{M_\odot}$}}
\newcommand{\msol}{\hbox{$\mathrm{M_\odot}$}}
\newcommand{\zfourge}{ZFOURGE}
\newcommand{\ZFOURGE}{ZFOURGE}

\newcommand{\myemail}{kawinwanichakij@physics.tamu.edu}
\begin{document}
\title{The Distribution of Satellites around Massive Galaxies at $1<z<3$ in ZFOURGE/CANDELS: Dependence on Star Formation Activity\footnote{T\lowercase{his paper includes data gathered with the 6.5 meter} M\lowercase{agellan} T\lowercase{elescopes located at} L\lowercase{as} C\lowercase{ampanas} O\lowercase{bservatory}, C\lowercase{hile}.}}
\author{Lalitwadee Kawinwanichakij\altaffilmark{1$\dagger$}}
\author{Casey	Papovich\altaffilmark{1}}
\author{Ryan F. Quadri\altaffilmark{1,2,3,4}}
\author{Kim-Vy H. Tran\altaffilmark{1}}
\author{Lee R.	Spitler\altaffilmark{5,6}}
\author{Glenn G. Kacprzak\altaffilmark{7,8}}
\author{Ivo Labb\'{e}\altaffilmark{9}}
\author{Caroline M. S. Straatman	\altaffilmark{9}}
\author{Karl Glazebrook\altaffilmark{ 7 }}
\author{Rebecca Allen\altaffilmark{6,7}}
\author{Michael Cowley\altaffilmark{5}}
\author{Romeel Dav\'{e}\altaffilmark{10,11,12}}
\author{Avishai Dekel\altaffilmark{13}}
\author{Henry C. Ferguson\altaffilmark{14}}
\author{W. G Hartley\altaffilmark{15,16}}
\author{Anton M. Koekemoer\altaffilmark{14}}
\author{David C. Koo\altaffilmark{17}}
\author{Yu	 Lu\altaffilmark{18}}
\author{Nicola	 Mehrtens\altaffilmark{1}}
\author{Themiya Nanayakkara\altaffilmark{7}}
\author{S. Eric	Persson\altaffilmark{3}}
\author{Glen Rees\altaffilmark{5}}
\author{Brett Salmon\altaffilmark{1}}
\author{Vithal	Tilvi\altaffilmark{1}}
\author{Adam R. Tomczak\altaffilmark{1}}
\author{Pieter 	van Dokkum\altaffilmark{19}}

\altaffiltext{1}	{George P. and Cynthia W. Mitchell Institute for Fundamental Physics and Astronomy, Department of Physics and Astronomy,Texas A\&M University, College Station, TX 77843}						
\altaffiltext{2}	{Mitchell Astronomy Fellow}						
\altaffiltext{3}{Carnegie Observatories, Pasadena, CA 91101, USA}						
\altaffiltext{4}{Hubble Fellow}						
\altaffiltext{5}{Department of Physics and Astronomy, Faculty of Sciences, Macquarie University, Sydney, NSW 2109, Australia}						
\altaffiltext{6}{Australian Astronomical Observatories, PO Box 915, North Ryde NSW 1670, Australia}						
\altaffiltext{7}{Centre for Astrophysics and Supercomputing, Swinburne University, Hawthorn, VIC 3122, Australia}						
\altaffiltext{8}{Australian Research Council Super Science Fellow}						
\altaffiltext{9}{Leiden Observatory, Leiden University, P.O. Box 9513, 2300 RA Leiden, The Netherlands}						
\altaffiltext{10	}{University of the Western Cape, Bellville, Cape Town 7535, South Africa}						
\altaffiltext{11	}{South African Astronomical Observatories, Observatory, Cape Town 7925, South Africa	}					
\altaffiltext{12	}{African Institute for Mathematical Sciences, Muizenberg, Cape Town 7945, South Africa}						
\altaffiltext{13}{Center for Astrophysics and Planetary Science, Racah Institute of Physics, The Hebrew University, Jerusalem 91904, Israel}						
\altaffiltext{14}{Space Telescope Science Institute, 3700 San Martin Drive, Baltimore, MD 21218, USA}						
\altaffiltext{15}{School of Physics and Astronomy, University of Nottingham, Nottingham NG7 2RD, UK}						
\altaffiltext{16}{Institute for Astronomy, ETH Zurich, Wolfgang-Pauli-Strasse 27, CH-8093 Zurich,Switzerland}					
\altaffiltext{17	}{University of California Observatories/Lick Observatory, Department of Astronomy and Astrophysics, University of California, Santa Cruz, CA 95064, USA}						
\altaffiltext{18	}{Kavli Institute for Particle Astrophysics \& Cosmology, 452 Lomita Mall, Stanford, CA 94305, USA	}					
\altaffiltext{19	}{Department of Astronomy, Yale University, New Haven, CT 06520, USA}		
\altaffiltext{$\dagger$}{\myemail}

\begin{abstract} 
We study the statistical distribution of satellites around star-forming and quiescent central galaxies at $1<z<3$ using imaging from the FourStar Galaxy Evolution Survey (\ZFOURGE) and the Cosmic Assembly Near-IR Deep Extragalactic Legacy Survey (CANDELS).  The deep near-IR data select satellites down to $\log(M/\msol)>9$ at $z<3$.  The radial satellite distribution around centrals is consistent with a projected NFW profile.  Massive quiescent centrals, $\log(M/\Msol)>10.78$, have $\sim$2 times the number of satellites compared to star-forming centrals with a significance of 2.7$\sigma$ even after accounting for differences in the centrals' stellar-mass distributions.  We find no statistical difference in the satellite distributions of intermediate-mass quiescent and star-forming centrals, $10.48<\log(M/\msol)<10.78$.  Comparing to the \citet{Guo2011} semi-analytic model, the excess number of satellites indicates that quiescent centrals have halo masses 0.3~dex larger than star-forming centrals, even when the stellar-mass distributions are fixed. We use a simple toy model that relates halo mass and quenching, which roughly reproduces the observed quenched fractions and the differences in halo mass between star-forming and quenched galaxies only if galaxies have a quenching probability that increases with halo mass from $\sim$0 for $\log(M_h/\msol)\sim$11 to $\sim$1 for $\log(M_h/\msol)\sim$13.5.  A single halo-mass quenching threshold is unable to reproduce the quiescent fraction and satellite distribution of centrals.   Therefore, while halo quenching may be an important mechanism, it is unlikely to be the only factor driving quenching.  It remains unclear why a high fraction of centrals remain star-forming even in relatively massive halos.
\end{abstract}


\section{Introduction}

One of the fundamental goals in studying galaxy formation and
evolution is to understand the relationship between galaxies and their
host dark matter halos. In $\Lambda$CDM models galaxies grow
hierarchically, and we expect to see the signatures of such growth in
satellite galaxies, which trace the accreted dark-matter
sub-halos. The distribution of satellite galaxies can be used as a
tracer of the overall mass distribution of the parent halo and
provides constraints on the halo mass. Therefore, investigating the
distribution of satellites provides a means to study how properties of
host galaxies (such as stellar mass and star formation activity) are
related to the mass of their dark matter halos.

\par  The distribution of satellite galaxies provides constraints 
that are independent of other techniques that
use correlation functions, galaxy-galaxy lensing, adundance matching,
and kinematics to study how the dark matter masses relate to galaxy observables
\citep[e.g.][]{Vader1991,Zehavi2002,Madore2004,Mandelbaum2006,Masjedi2006,
Gavazzi2007,Wake2008,Conroy2009,Drory2009,Behroozi2010,Hartley2010,
Hartley2013,More2011}. Several studies show that the
distribution of satellite galaxies  follows a Navarro, Frenk, \& White
(1996) profile (NFW-profile), and this can be used to measure the
mass distribution and scale size of the dark matter halo
\citep{Nierenberg2011,Nierenberg2012,Tal2012,Tal2013,Wang2012,Watson2012,Wang2014}.  

\par Studying the distribution of satellite galaxies is expected to
provide particular insight  into galaxy evolution for massive
galaxies. Growing observational  evidence suggests that massive ($>
10^{11 }~\msol$), quiescent galaxies (those lacking substantial
star-formation) have grown primarily in their outer regions through
the accretion of small satellites since $z\sim2$
\citep[e.g.][]{VanDokkum2010,McLure2012,Greene2012,Greene2013}. This
is  consistent with the results from hydrodynamical simulations, which
have reproduced the observed size and mass growth of massive
elliptical galaxies by stellar accretion from minor mergers
\citep{Naab2007,Naab2009,Oser2010,Oser2012,Hilz2013}. 

\citet{Tal2012} use a statistical background subtraction to
measure the radial number-density profiles of satellites around local
luminous red galaxies (LRGs). They show that the best fit
NFW+S\'{e}rsic model of the derived satellites profile results in a
total dark-to-baryonic mass ratio in agreement with the weak-lensing
result from \citet{Mandelbaum2006} for massive early type
galaxies. Tal et al.\ (2013) extend this technique to massive galaxies
out to $z\sim 2$ and show that the radial number density profile of
satellite galaxies has not evolved significantly since $z = 1.6$,
suggesting a balance between mergers and accretion of new satellites.

Recently, \citet{Hartley2014} have used a similar method to study the properties of satellites of centrals out to $z < 1.9$ using data in the UKIDSS UDS, albeit with a higher mass limit for satellites, $\log M/M_\odot > 9.7$.

The technique used by \citet{Tal2012,Tal2013} requires  statistically
isolating satellites from unassociated galaxies along the line of
sight.  Surveys with homogeneous multiwavelength photometry provide
photometric redshift information on faint galaxies, which is useful to
identify satellites around more distant galaxies as shown by Tal et al
(2013). This statistical technique has an advantage over methods using
spectroscopic redshifts, as the latter are observationally prohibitive
for all but the brightest satellites and very costly in telescope
time.  

\par At $z > 1$, growing evidence shows that massive galaxies are a
mix of quiescent and star-forming populations \citep[e.g.][]{Papovich2006}.  This is an extension of the well-known bimodality in their
color and star-formation activity.   At fixed stellar mass, galaxies
divide into star-forming galaxies with bluer UV--optical colors and
active star formation \citep[forming a ``main sequence'',
e.g.,][]{Noeske2007,Martin2007},  and quiescent galaxies with red
colors and low star formation rates compared to their past average
\citep[forming a ``red sequence'', e.g.][]{Bell2004,Blanton2006}.
This bimodality in the galaxy color-mass distribution extends at least
out to $z\sim 3$ \citep{Whitaker2011,Tomczak2014} and perhaps beyond
\citep[Spitler et al. 2014,][]{Straatman2014}.  Recent studies of the
evolution of the progenitors of local ultra-massive galaxies
($\log{(M/\Msol)}\approx11.8$) at $1 < z < 3$ shows that the
contribution of star-forming galaxies increases at $z >
1$. Understanding why some massive galaxies are quiescent and some are
star-forming has important implications for galaxy formation models
\citep[e.g.][]{Croton2006,Martig2009,Dekel2006}. This may be tied to
differences in the galaxies' dark matter halo growth. For this reason,
it is of interest to study differences in the satellite distribution
of quiescent and star-forming centrals at $1 < z < 3$ because it
allows us to trace the dark-matter halos of such objects when the
dichotomy in star-formation activity (quiescent versus star-forming)
is at its peak for massive galaxies \citep[e.g.][]{Marchesini2014}   

\par In this work, we use the most recent data from the FourStar
Galaxy Evolution Survey (ZFOURGE) and the Cosmic Assembly Near-IR Deep
Extragalactic Legacy Survey (CANDELS) to derive the distribution of
satellites around massive galaxies at $1 < z < 3$. We study the
dependence of the galaxy distribution on the star formation activity
of the massive centrals.   The outline of this paper is as follows.
In \S~2 we describe our \zfourge/CANDELS dataset and our galaxy sample
selection criteria.   In \S~3 we describe the method for identifying
satellites and for measuring the satellite number density profile.  In
\S~4 we discuss how the satellite distribution depends on the stellar
mass  and star-formation activity of central galaxy.  In \S~5, we
explore why quiescent and star-forming galaxies have differences in
their satellite distributions, including a comparison between galaxies
in our dataset and those in a semi-analytical model \citep{Guo2011}.
In \S~6, we present our summary.  Throughout, we adopt the following
cosmological parameters where appropriate, $H_0 = 70
\unit{km\;s^{-1}\;Mpc^{-1}}$, $\Omega_{m} =0.3$, and $\Omega_{\Lambda}
= 0.7$.   For this cosmology, the angular diameter conversion is
$\approx$ 8 kpc arcsec$^{-1}$ and constant within 5\% for $1 < z <
3$. 

\section{Data and Sample Selection}
\label{sec:data}

 We use the deep near-IR imaging from ZFOURGE (Straatman et al. in
prep\footnote{http://zfourge.tamu.edu}), which is composed of three $11\arcmin \times 11\arcmin$
pointings with coverage in the CDFS \citep{Giacconi2002}, COSMOS
\citep{Capak2007}, and UDS \citep{Lawrence2007}. The imaging reaches
depths of $\sim 26 \unit{mag}$ in $J_{1}, J_{2}, J_{3}$ and $\sim 25
\unit{mag}$ in $H_{s}, H_{l}, K_{s}$. A brief description of the
filter set is described in detail by \citet{Spitler2012,Tilvi2013}.
The medium-band filters from ZFOURGE provide an advantage by sampling
the Balmer break at $1<z<4$ better than broadband filters alone.  We
combine the ZFOURGE data with public HST/WFC3 F160W and F125W imaging
from CANDELS \citep{Grogin2011,Koekemoer2011} in the three fields. As
described in \cite{Tomczak2014}, we make use of the CANDELS F160W as a
detection image to preselect a sample of galaxies at $z<3$ to low
masses ($10^{9}~\msol$). 

Photometry is performed in dual-image mode with SExtractor
\citep{Bertin1996} on PSF-matched images. The colors are measured in
0.8\arcsec apertures, and total magnitudes are measured using the AUTO
magnitude and applying an aperture correction for the flux falling
outside the AUTO aperture.

As described by \cite{Tomczak2014}, we estimate the photometric
redshifts and rest-frame colors of galaxies with EAZY
\citep{Brammer2008}. By using the default set of spectral
templates derived from the PEGASE models \citep{Fioc1997} and a dust
reddened template derived from the \cite{Maraston2005} model to fit
the $0.3-8\ \mu\unit{m}$ photometry for each galaxy to obtain its
photometric redshift.  Similarly, we derive stellar masses using
\cite{Bruzual2003} stellar population models with FAST code
\citep{Kriek2009}, assuming exponentially declining star formation
histories, solar metallicity, and a \cite{Chabrier2003} initial mass
function.

We estimate relative uncertainties in photometric redshifts
using the technique described by \citet{Quadri2010}.   For our study,
the relative errors between the centrals and satellites are
paramount, and traditional photometric redshift testing (comparing
photometric redshifts to spectroscopic) is infeasible as the satellite
galaxies in our sample are typically much fainter than spectroscopic
magnitude limits.   The underlying principle of the Quadri \& Williams
technique is that close pairs of galaxies have some probability of being
physically associated.  In this case, each galaxy provides an independent 
estimate of the true redshift.  Therefore, the distribution of the
differences in the photometric redshifts of galaxy pairs can be used
to estimate the photometric redshift uncertainties. 

We apply this method to derive the distribution of differences between
the photometric redshifts of centrals and satellites using the samples
defined below (\S~2.1 and \S~2.2).  From these we find that the
typical photometric redshift uncertainty between the centrals and
satellites in  the COSMOS, CDFS, and UDS fields are $\sigma_z$= 0.06,
0.05, and 0.08, respectively (where $\sigma_z =\sigma /\sqrt{2}$, and
where $\sigma$ is the width measured from a Gaussian fit to the
distribution of pair redshift differences in each field, and the
$\sqrt{2}$ accounts for the fact that we take the difference between
two independent measurements).  These uncertainties translate to
$\sigma_z / (1+z) < 2$\% and 4\% for galaxies at $1 < z < 3$ down to
$10^9$~\msol. 

Throughout, we consider two samples of galaxies, the central galaxies
and their satellite galaxies, which are defined in \S~2.1 and
\S~2.2. We denote the stellar masses of the centrals as $M_\mathrm{c}$
and the stellar masses of the satellites as $M_{\mathrm{sat}}$. We use
$N_{\mathrm{sat}}$ throughout to denote the radial number density of
satellites around the central galaxies, and correct this quantity for
projected field galaxies (\S~3.1). 

\begin{deluxetable}{cccc}
\tabletypesize{\footnotesize}
\tablecolumns{4} 
\tablewidth{0pt}
 \tablecaption{ Number and mean stellar masses of quiescent centrals and star-forming centrals in ZFOURGE survey at $1 < z < 3$
 \label{table:sampleno}}
 \tablehead{
 \colhead{Field} \vspace{-0.2cm}& \colhead{$\log(M_{\unit{c}}/\Msol)$} &  \colhead{$N_{\unit{c}}$} & \colhead{$\log(M_{\unit{c,mean}}/\Msol$)}\\ 
\vspace{0.1cm}} 
 \startdata 
\vspace{0.2cm}
  
  Quiescent centrals &      &     & \\ 

       COSMOS  &      10.48--10.78 & 63 & 10.61 \\ 
                     &       $> 10.78$    & 67 & 11.01 \\ 
      CDFS             &      10.48--10.78 & 46 & 10.61\\ 
                     &       $> 10.78$    & 53  & 11.02 \\ 
     UDS              &      10.48--10.78 & 96 & 10.61 \\
                      &       $> 10.78$    & 70  & 11.04  \\ 
 
      Star-forming centrals &      &  & \\ 
      & & \\
       COSMOS &      10.48--10.78 & 85 & 10.62 \\
                     &       $> 10.78$    & 71 & 11.08   \\ 
       CDFS      &      10.48--10.78 &83 & 10.62  \\ 
                     &       $> 10.78$    & 52  &11.03 \\ 
       UDS        &      10.48--10.78 & 87 & 10.62\\
                     &       $> 10.78$    & 68  & 10.97\\ 
 \enddata
\vspace{-0.3cm}

\end{deluxetable}

\subsection{Selection of Centrals}
\label{sec:samplesel}
Our goal is to measure the distribution of satellites around massive
galaxies at $1 < z < 3$. We therefore select all galaxies in \ZFOURGE\
with $\log(M_{\mathrm{c}}/\Msol) > 10.48$ (i.e., $M_\mathrm{c}  > 3
\times 10^{10}$~\msol) and photometric redshift $1 < z < 3$ as our
sample of central galaxies.   We will further consider the subsamples of
central galaxies in bins of stellar mass, $10.48 <
\log (M_\mathrm{c}/\msol) < 10.78$ (i.e., $M_{\mathrm{c}} = (3-6) \times
10^{10}$~\msol) and $\log (M_\mathrm{c}/\msol) > 10.78$ (i.e.,
$M_\mathrm{c} > 6 \times 10^{10}$~\msol).   A summary of number and 
mean stellar mass of centrals in each of the \ZFOURGE\ field and in each subsample is
given in Table ~\ref{table:sampleno}. 

\begin{deluxetable*}{ccccc}
\tabletypesize{\footnotesize}
\tablecolumns{5} 
\tablewidth{0pt}
 \tablecaption{ Summary of the probabilities $p$-values comparing the number density of satellites around quiescent and star-forming centrals at $1 < z < 3$ in the ZFOURGE survey.  
 \label{table:pvaltable}}
\tablehead{
 \colhead{Sample} \vspace{-0.2cm} &\colhead{$\log(M_{\unit{c}}/\Msol)$} &\colhead{$N_{\unit{c}}(\mathrm{Quiescent})$} & \colhead{$N_{\unit{c}}(\mathrm{Star-forming})$} &\colhead{$p_{\mathrm{MC}}$} \\ \\
\colhead{(1)} &\colhead{(2)} &\colhead{(3)} &\colhead{(4)} &\colhead{(5)}  \\
 \vspace{0.1cm}} 
 \startdata 
\vspace{0.1cm}

All centrals  &  $> 10.48$ & 395 & 446 & 0.081 \\
                     &  10.48-10.78 & 205 & 255  & 0.478 \\
                     &  $> 10.78$ & 190 & 191  & 0.002 \\

Fixed in stellar mass & $ > 10.48$ & 337 & 337 & 0.550 \\
&  10.48-10.78 & 195 & 195 & 0.985 \\
 &  $> 10.78$ & 142 & 142 & 0.004 \\
\enddata
\vspace{-0.3cm}
  \tablecomments{(1) Description of samples used, (2) stellar mass of the sample, (3) number of quiescent centrals, (4) number of star-forming centrals, and (5) The probability we derive from our Monte Carlo simulations that we would have obtain a difference between the satellite distributions of the quiescent and star-forming centrals by chance. Low probabilities ($p$-value) indicate more significant differences in the distributions.}
\end{deluxetable*}

\begin{figure}[h]
\epsscale{1.1}
\plotone{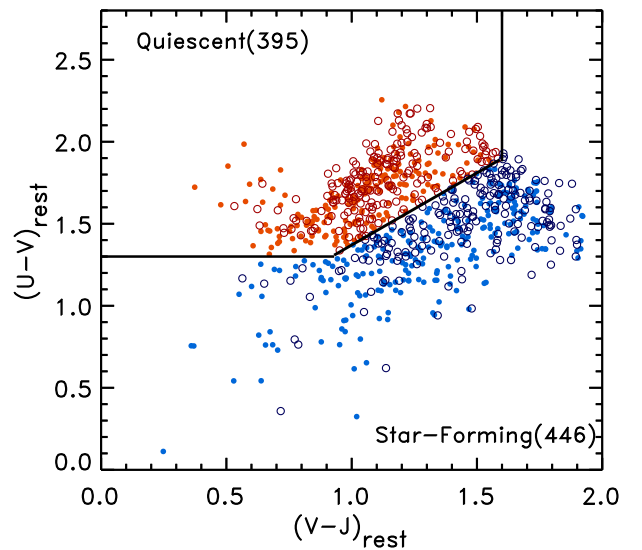}
\caption{Rest-frame $U-V$ versus $V-J$ color for massive central
galaxies in our sample from all three ZFOURGE pointings 
(COSMOS, CDFS, and UDS) at $1 < z < 3$.  The data points show centrals with
stellar masses of $10.48 <\log(M_{\unit{c}}/\Msol) < 10.78$ (small
filled circles) and  $\log(M_{\unit{c}}/\Msol) > 10.78$ (large
open circles). The galaxies in the upper left region of the plot
(separated by the solid line) are quiescent (red open and
filled circles); galaxies outside this region are star forming 
(blue open and filled circles), using the definition of
\cite{Williams2009} and Equation~1.  The numbers of quiescent and
star-forming centrals are indicated in parentheses in the plot
legend. Quiescent and star-forming galaxies each account for
approximately half of galaxies population at these masses and
redshifts.}
\label{fig:uvjdiagram}
\end{figure}

According to the central galaxies selection criteria used by
\citet{Tal2013},  they consider galaxies as ``central''  if no other,
more massive galaxies are found within a projected radius of 500
kpc. Otherwise, they are counted as “satellites” of their more massive
neighbor.   We have tested if the projected radial distribution
changes if we exclude galaxies from our sample of central galaxies
if there are  other more massive galaxies within 10 arcsec (about 80
kpc, projected), which is comparable to our derived halo
scale radius (described in \S~3.2, below).  We find that our derived
projected radial  distribution is not significantly
changed. Therefore, we do not apply this isolation criteria to
select our central sample. However, this may introduce
  galaxies which are satellites into our sample of centrals.  In
  \S~5.2 we further quantify the effects of the misclassification of
  centrals and satellites on the number density of satellites using
  results from a mock galaxy catalog.

We then consider the subsamples of central galaxies divided by
star-formation activity.  \citet{Williams2009} show that the $U-V$ and
$V-J$ rest-frame colors are able to distinguish reliably between
quiescent galaxies with low specific SFRs (sSFR) and star-forming
galaxies with  high specific SFRs; see also discussion in
\citet{Whitaker2011}. Galaxies are classified as quiescent if their
rest-frame colors satisfy these criteria:
\begin{eqnarray}
U-V&>& 0.88 \times (V-J) +0.49      \nonumber \\
   U-V&>& 1.3  \\
   V-J&<&1.6 \nonumber
\end{eqnarray}
Figure~\ref{fig:uvjdiagram} shows the $U-V$ vs. $V-J$ diagram
(hereafter $UVJ$ diagram) for the centrals in our \zfourge\ samples.
We find that between $1 < z < 3$ the massive centrals (841 in total
with $M_\mathrm{c} > 3 \times 10^{10}$~\msol) are roughly evenly
divided into quiescent galaxies (47\%) and star-forming galaxies
(53\%) based on their $UVJ$ colors.  The numbers of the quiescent and
star-forming centrals in each \ZFOURGE\ field and mass subsample are
shown in Table ~\ref{table:sampleno}. 

\subsection{Selection of Satellites}
\label{sec:selsat}

To identify satellites of the central galaxies in our sample, we build
on the statistical background subtraction technique, as discussed in
\cite{Tal2012,Tal2013}.   We first select all galaxies around each
central from our ZFOURGE catalogs that satisfy the following
conditions
\begin{eqnarray}
\left| z_\mathrm{c} - z_\mathrm{sat} \right| \leq 0.2 \nonumber \\
10^9\ \msol \leq M_\mathrm{sat} < M_\mathrm{c} \\
\end{eqnarray}
where $z_\mathrm{c}$ and $z_\mathrm{sat}$ are the
photometric redshift of the central and satellite, respectively.
Similarly, $M_\mathrm{c}$ and $M_\mathrm{sat}$ are the stellar mass of
the central and satellite, respectively.   Our requirement that
$\Delta z = |z_\mathrm{c} - z_\mathrm{sat}| \leq 0.2$ is motivated by
our relative photometric uncertainty ($\sigma_z$) between centrals
and satellites derived above.  In each case, the $\sigma_z$ values for
galaxies in each \zfourge\ field are less than about half the $\Delta
z \leq 0.2$ requirement in Equation~2, which argues that this
selection criterion is appropriate.

The mass-completeness limits for all galaxies in the \zfourge\ sample
at $z=3$ are $\log(M/\Msol) = 9.3$ \citep{Tomczak2014}. Below these
mass limits, we are incomplete for quiescent galaxies, while our
sample remains complete for star-forming galaxies down to $\log
(M/\Msol) = 9$. However, in this study, we are comparing the relative
number of satellites between quiescent and star-forming centrals, so
they have the same relative bias due to incompleteness in satellite
detection. Parenthetically, we note that we have also
  repeated our analysis restricting the centrals to  $1 < z <
2$, where incompleteness is less of an issue, and our primary
conclusions are unchanged.

We consider the possibility that our samples of quiescent and
star-forming centrals have different redshift distributions.  For
example, it could be plausible that the quiescent centrals tend to
have lower redshifts, while star-forming centrals tend to have higher
redshift. In this hypothetical case, we might expect more satellites
around quiescent galaxies because we can see them to lower mass. We
therefore compared the redshift distributions of the different
subsampltes. For the intermediate mass star-forming and
quiescent centrals ($10.48 < \log(M_{\mathrm{c}}/\Msol) < 10.78 $),
there is no difference in their redshift distributions. There is some
difference in the redshift distributions between the star-forming and
quiescent centrals in the higher mass subsample.  However, in \S~5.1 we
show that this difference does not effect our main results. Therefore,
we conclude that the redshift distributions of central galaxies does
not affect the relative number of satellites between the star-forming
and quiescent samples.

\section{Radial Number Density Profiles}
\subsection{Profile derivation}
\label{sec:profilederive}

\begin{figure*}
\epsscale{1.0}
\plotone{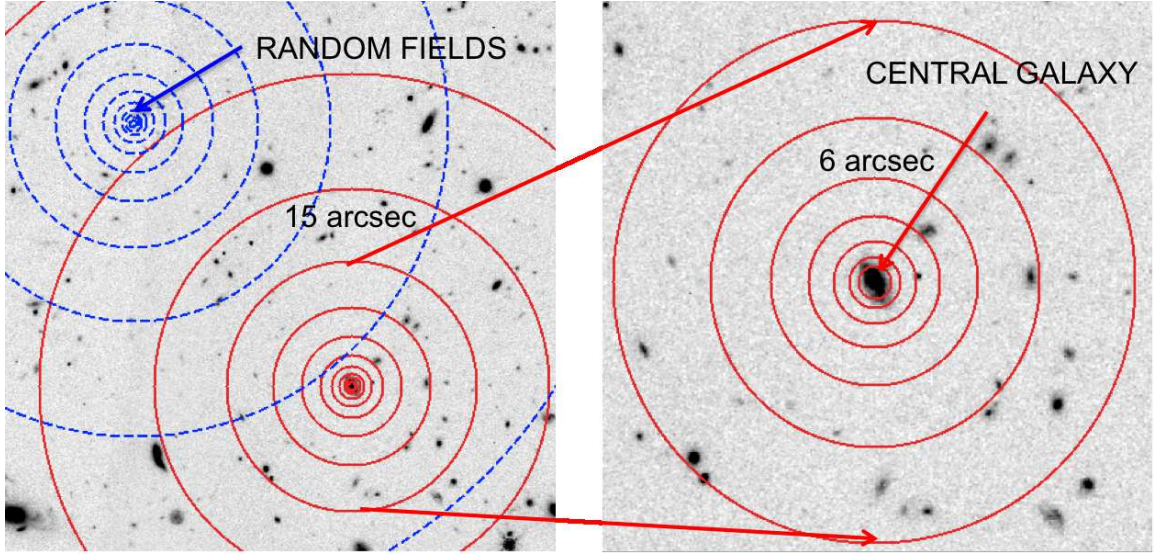}

\caption{Demonstration of the statistical technique to measure the
radial number density profile of satellite galaxies.  All galaxies
that satisfy the definition of ``satellite'' in Equation~2 are divided
into log($r$) bins around each central (red solid circles) and also
around randomly selected positions in the field (blue dashed
circles). The number density profile of satellites is then measured
from the difference between the satellites measured around the central
and those measured in the field. In practice, we use many random
fields per each central and calculate the average to infer the
statistical distribution of foreground and background galaxies. In
this figure we show one random field for illustration only.}
\label{fig:demons}
\end{figure*}

\begin{figure*}
\epsscale{1.0}
\plotone{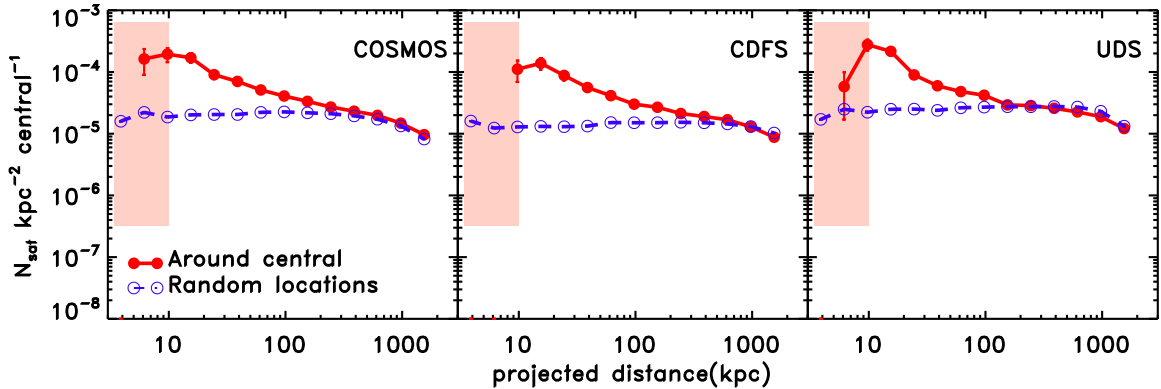}
%
\caption{The average projected radial profile of satellites around
central galaxies at $1 < z < 3$ with stellar masses of
$\log(M_{\mathrm{c}}/\Msol)> 10.48 $ and the field in each pointing of
\ZFOURGE\ (COSMOS, CDFS, and UDS). The measured projected distribution
of satellites is calculated by  subtracting the average random profile
(blue lines with open circles) from that measured around the central
galaxy profile (red lines with filled circles). The shaded areas show
the range of projected distance where satellites are blended with
centrals; these we exclude from our analysis. Uncertainties on the
random profiles are small compared to the symbol sizes, and are not
shown.}
\label{fig:nrealnrand}
\end{figure*}

\begin{figure*}
\epsscale{1.0}
\plotone{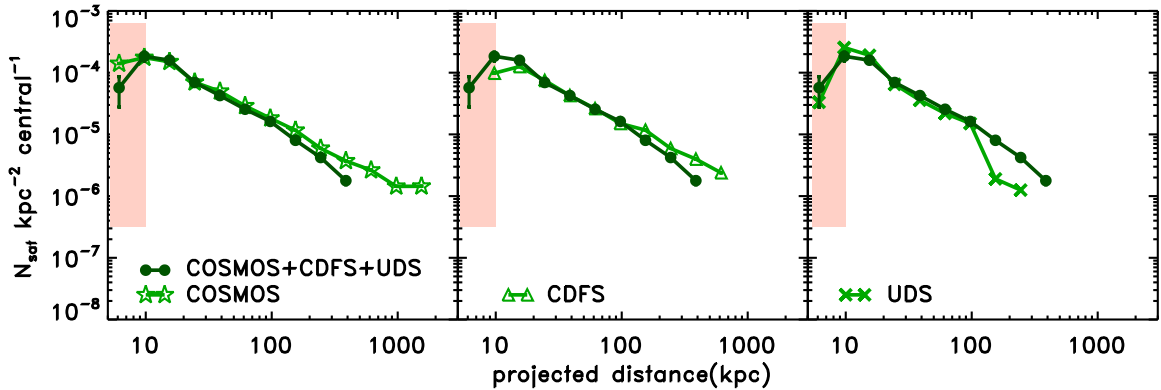}

\caption{The projected radial profile for central galaxies at $1 < z <
3$ with stellar masses of $\log(M_{\unit{c}}/\msol)  > 10.48$ combined
from all \ZFOURGE\ fields (dark green solid line),  compared to the
profiles of each field separately (light green).  In each panel, the
combined measurement is the same, where each field shows the result
from COSMOS (left panel), CDFS (middle panel), and  UDS (right
panel). The profile of each field separately is consistent with the
combined one at $10 < r/\mathrm{kpc} < 100$, but differences exist at
larger projected radii, which we attribute to variations in the
background of field galaxies. } 

\label{fig:ffvariation}
\end{figure*}

\par The method to extract a radial distribution of satellites is
illustrated in Figure~\ref{fig:demons}, expanding from the method
outlined in Tal et al.\ (2012). For a given projected distance from
each central, we measure the number of all galaxies satisfying our
definition of satellite in Equation~2.  This includes both physically
associated galaxies, as well as chance alignments of foreground and
background galaxies. We measure the projected radial distribution by
binning the distance from those galaxies to central galaxies in
logarithmic bins. 

\par We remove the contamination from foreground and background
galaxies statistically by repeating the measurements in randomly
selected positions within the entire area of the \ZFOURGE\
fields. While we use all centrals, some centrals  are near the
survey edges, which restricts our ability to detect their
satellites. To account for this, we correct the galaxy counts in each
annulus by the fraction of the annulus that falls outside of the
survey boundaries. The locations of random annuli, and the
corrections, are determined in the same way as for the central
galaxies. The only other restriction we place on the random fields
is that they are not centered within 6 arcsec (about 50 kpc,
projected) of any central.  We have not
required that the random fields have zero overlap with areas around
our centrals, as the surface density of our centrals is $\sim 2
\unit{arcmin}^{-2}$, and such a constraint would be too prohibitive.
This is a tradeoff between our requirement to subtract off
statistically the foreground and background galaxies, and having a
sufficiently high number of the random fields to measure the
background accurately in each \zfourge\ field.  We then subtract the
number density of satellites in the random pointings  from the number
density of satellites of each central. In practice, we measure the
number density of galaxies in 100 random pointings and take
the average to estimate the number density of foreground and
background galaxies for each central in each \zfourge\ field. 

Figure~\ref{fig:nrealnrand} shows the raw number density of galaxies
measured around both the centrals and measured in random pointings.
In each field there is a strong statistical excess of galaxies around
our centrals extending from ~10 kpc to $\sim$ 100--400 kpc
($\sim$12--50 arcsec), which we attribute to physically
associated satellites. There are slight variations in the
$N_{\mathrm{sat}}$ distribution inferred from the background, as we
would expect from natural field-to-field variations.  We see no
substantial variation in the $N_{\mathrm{sat}}$ projected
distributions between the three separate fields.

The restriction on the location of the random aperture has little
effect on our conclusions. We have tested if this signal changes if we require that no random background
aperture is  centered within 12 arcsec of a central (compared to the 6
arcsec requirement above), but we find that this does not change
significantly the number density of galaxies in the random
pointings, and therefore this does not affect our measurement. Similarly, we also find that there is no significant change if we place no
restrictions on the locations of the random background apertures.

At smaller projected distances ($<$ 10 $\mathrm{kpc}$), we are unable
to measure reliably the number density of satellites, as such objects
are blended with the isophotes of the central galaxies.   For example,
we cross-match our central galaxy catalog from \ZFOURGE\ to the
morphology parameter  catalog for the CANDELS WFC3/F160W imaging from
\citet{VanderWel2012}.  From this, the typical effective radii of our
centrals at $1 < z < 3$ is $\sim 3.2 \unit{kpc}$, consistent with
measurements of massive galaxies at $z\sim 2$ from
\cite{VanDokkum2010}.   Furthermore, van Dokkum et al.\ show that such
galaxies have $\sim 1 \%$ of their stellar mass at a distance of $10
\unit{kpc}$.  Therefore, it seems likely that satellites around  these
galaxies would be indistinguishable from substructure in the centrals
for projected distances $r < 10$~kpc.  Indeed, doing a careful
analysis by subtracting the light from the central, Tal et al.\ (2012)
find that the number density of satellites around centrals follows a
$r^{1/4}$-law, consistent with the surface-brightness profile of the
central galaxies. This suggests that radial density profile of
satellites  at small scales is strongly influenced by the baryonic
content of the central galaxy, rather than the dark matter halo.

Figure~\ref{fig:ffvariation} shows the number density of satellites
measured for the centrals in each \zfourge\ field, and for the
combined sample.   The satellite distribution in each field is
consistent with that measured in the combined sample for $10 <
r/\unit{kpc} < 100$, and we observe differences at larger and
smaller radii.  \ZFOURGE\ contains three largely separated fields on the
sky,  and we interpret the differences at larger radii as a
result of field-to-field variations in the number density of
background/foreground galaxies in each field. 

\begin{figure}
\epsscale{1.0}
\plotone{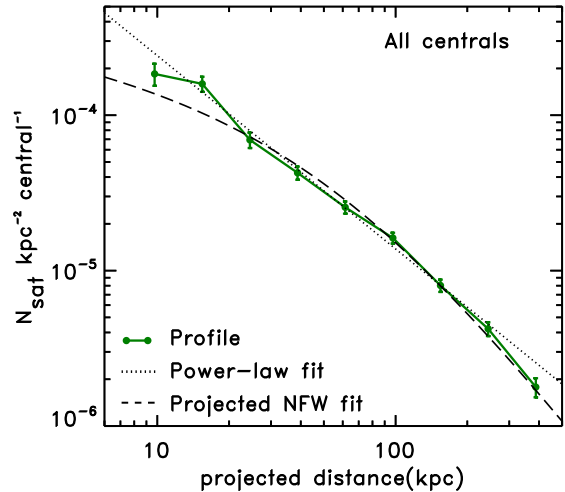}

\caption{The average projected radial profile of satellites around all
central galaxies at $1 < z < 3$ with stellar masses of
$\log(M_{\unit{c}}/\Msol)  > 10.48$ (green solid line) combined from
all \zfourge\ fields.  The radial profile is  fitted well with a
power-law model with $r^{-1.2}$ (dotted line) and a projected
NFW model with $r_s=61.1 \pm 7.8$ kpc (dashed line) over the
range $10 < r/\unit{kpc} <350$.}

\label{fig:fitpowerlawnfw}
\end{figure}

\begin{figure*}
\epsscale{1.0}
\plotone{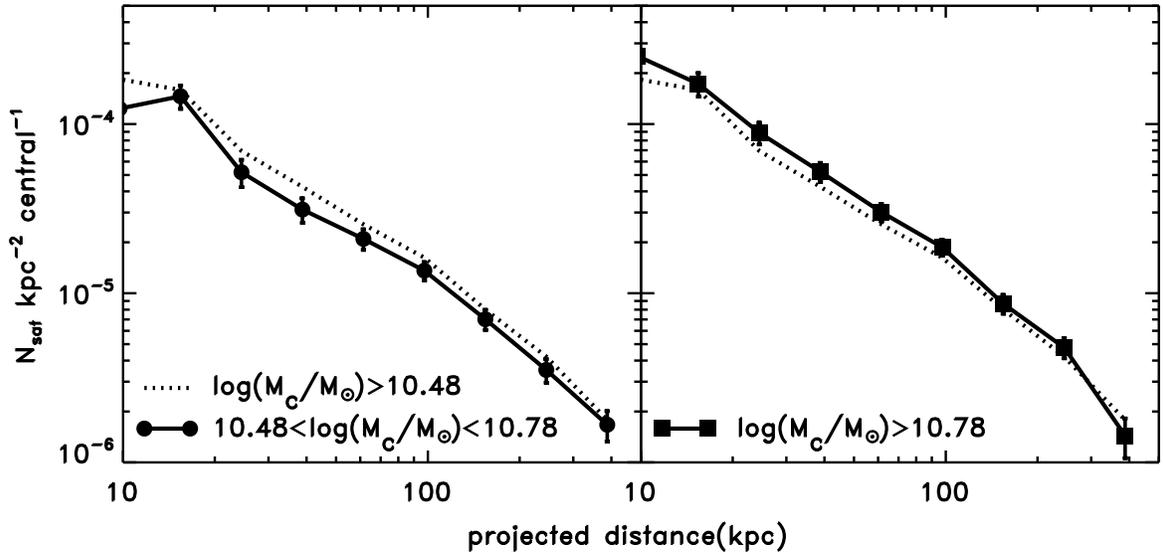}

\caption{The dependence of the number density of satellites on stellar
mass of the central.  \emph{Left:} the projected radial profile around
all ZFOURGE centrals at $1 < z < 3$ with stellar masses of $10.48 <
\log(M_{\unit{c}}/\Msol)  < 10.78$ (solid line with circles) compared
with that of all samples with stellar masses of
$\log(M_{\unit{c}}/\msol)  > 10.48$ (dotted line),   \emph{Right:}
same as the left panel but for the comparison between the profile of
satellites around centrals with stellar masses of
$\log(M_{\unit{c}}/\msol)  > 10.78$ (solid line with boxes) and the
profile of total centrals. The number of satellites around centrals
with stellar masses of $\log(M_{\unit{c}}/\msol)  > 10.78$ is higher
on average than that around centrals with stellar masses of $10.48 <
\log(M_{\unit{c}}/\msol)  < 10.78$ at $1.9\sigma$. }
\label{fig:masscompareunmatched}
\end{figure*}

\subsection{Model Fitting}
\label{sec:modelfit}

We fit the combined projected number-density profile of satellites
using a simple powerlaw model, where $N_\mathrm{sat} \propto
r^{\gamma}$. Figure~\ref{fig:fitpowerlawnfw} shows that in the range
$10 < r/\unit{kpc} < 350$ the profile is well  described by
the power-law with $\gamma = -1.24 \pm 0.04$.  The power-law
slope of the projected radial profile of satellites around luminous
red galaxies at $z=0.34$ is $-1.1$ \citep{Tal2012}, which is
marginally consistent with our measurement here. 

\par We also compare our measured projected radial profile of satellites
with a projected NFW profile, which \citet{Bartelmann1996} show is
$$
\Sigma(x) = \left\{ \begin{array}{ll}
 (x^2-1)^{-1}\left(1-\frac{2}{\sqrt{x^2-1}} \arctan\sqrt{\frac{x-1}{x+1}}\right) &\mbox{ $(x>1)$} \\
  1/3 &\mbox{$(x=1)$} \\
 (x^2-1)^{-1}\left(1-\frac{2}{\sqrt{1-x^2}} \mathrm{arctanh}\sqrt{\frac{1-x}{1+x}}\right) &\mbox{$(x<1)$} \\ 
       \end{array} \right.
$$
where $x \equiv r/r_{s}$ and $r_{s}$ is the NFW-profile scale
radius. We utilize the nonlinear least-squares curve fitting program
MPFIT \citep{Markwardt2009} to fit the projected NFW model to the
measured projected radial profile of satellites around centrals at
$10 < r/\unit{kpc} < 350$, where we fit for both the normalization
factor and scale radius. 
%
%
\par Figure~\ref{fig:fitpowerlawnfw} shows that our derived projected
radial density of satellites at $1<z<3$  is well fitted by the
projected NFW model with $r_s = 61.07 \pm 7.8 \unit{kpc}$.
Assuming the scale radius is independent of redshift and only scales
with a halo mass as $r_s \sim M_h^{0.45}$
\citep[e.g.,][]{Bullock2001}, this would predict $r_s
\simeq200$ ~kpc at $z=0$ (assuming a factor $\sim$10 growth in halo
mass, see e.g.,\citet{Moster2013}). This is smaller than that found by
\citet{Tal2012}, who find $r_s \sim 270$~kpc for galaxies at $z=0.02$
in SDSS, but the results are probably consistent as the galaxies in
their sample  correspond to progenitors with higher stellar masses by
factors of $\sim$3--5 compared to our sample here
\citep[e.g.,][]{Tal2013}.   Furthermore, this provides us with
confidence that our measured satellite distribution is tracing the
dark-matter halo of the centrals in our \zfourge\ samples.

\section{The Satellite Distribution Dependence on Galaxy Property}

\subsection{Significance estimation} \label{sec:errorestimate}

 It is desirable to assign a significance statistic ($p$-value) when
comparing the differences between the satellite number densities for
different subsamples. The uncertainties of each datum of our projected
radial distribution of satellites ($N_\mathrm{sat}$ in the figures
above) are derived using simple Poisson statistics.   When comparing
the satellite number density distributions for difference subsamples,
we use two methods, a direct rank-sum test and a Monte Carlo
simulation.   In practice, for reasons described below we find that
the Monte Carlo simulation provides more physical probabilities, and
we will use those to estimate the significance in our results.  

We first apply a one-sided Wilcoxon-Mann-Whitney rank-sum (WMW) test
 \citep{Mann1947} to quantify a probability that the number density of 
satellites around quiescent and star-forming centrals have the same 
parent population.  The WMW test measures a probability ($p$-value) 
using the data and Poisson errors on the satellite distributions between 
two subsamples (e.g., the quiescent and star-forming centrals).   
However, the WMW test is not strictly appropriate for our analysis 
because we are applying it to heavily binned data:  each datum is binned
(logarithmically) in radius over $10 < r / \mathrm{kpc} < 200$.     
In particular, the WMW test is insensitive to the fact that our sample includes 
hundreds of central galaxies and thousands of satellite galaxies.


To estimate meaningful $p$-values, we use a Monte
Carlo approach.  We create 10,000 simulations for each subsample of
central galaxies.  For a given stellar mass range of the central
subsample, we randomly select new samples of centrals from the
subsample (allowing replacement).  We then randomly assign each galaxy to be
either quiescent or star-forming.   In each simulated subsample, the
number of the quiescent and star-forming centrals are equal to the
actual number of each in the real subsamples.  We then recalculate
the radial number density of satellites for each set of random samples
and calculate the $p_\mathrm{WMW}$--value using the WMW test for each
iteration. The likelihood from the Monte Carlo simulations
$(p_{\mathrm{MC}})$ are calculated by determining the fraction of the
number of simulations when we have the $p_{\mathrm{WMW}}$ less than
the $p_\mathrm{WMW}$ value we derive from the real data.   A summary
of these likelihoods is given in Table ~\ref{table:pvaltable}.

\subsection{Dependence on Central Stellar Mass}

We study how the number density of satellites depends on the stellar
mass of centrals by dividing our central galaxy sample into two mass
bins: $10.48 < \log(M_{\unit{c}}/\Msol) < 10.78$ and
$\log(M_{\unit{c}}/\Msol) > 10.78$.  We then recompute the number
density of satellites for each of these subsamples using the method
above.  Figure~\ref{fig:masscompareunmatched} shows that the more
massive centrals have a higher number density of satellites compared
to the lower mass centrals.  Using our Monte Carlo simulations (see
\S ~\ref{sec:errorestimate}), we find that the significance of this result is 
$p_\mathrm{MC} = 0.026$ ($\simeq$1.9$\sigma$). Therefore, 
there is suggestive evidence that the number of satellites increases with the stellar 
mass of the central galaxy. 
\begin{figure*}
\epsscale{1.0}
\plotone{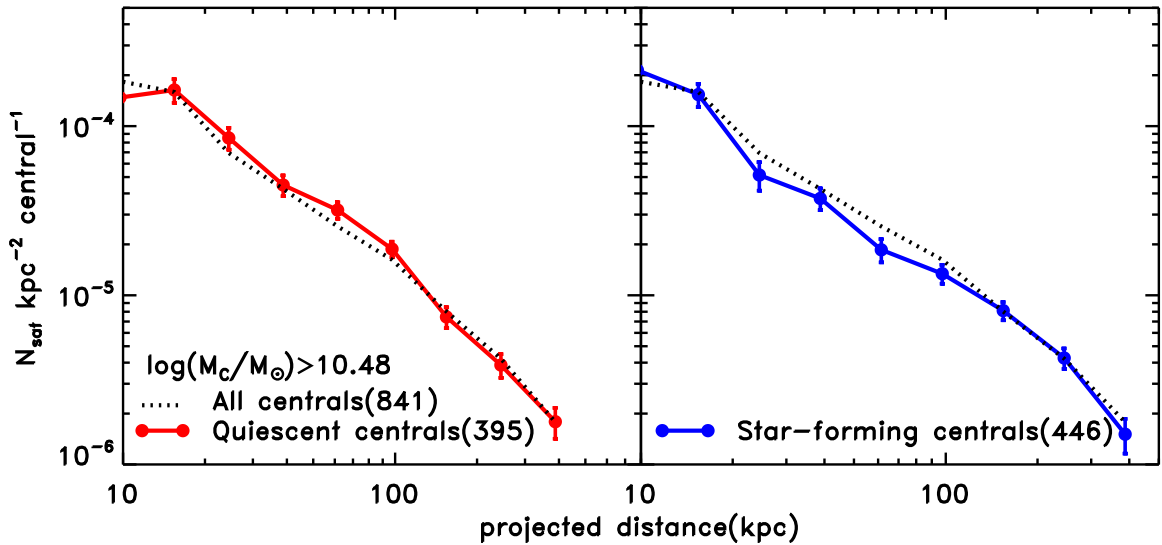}
%
\caption{The dependence of the number density of satellites on
star-formation activity of the central.  \emph{Left:} the projected
radial profile around all centrals (black dotted line) at $1 < z <3$
with stellar masses of $\log(M_{\unit{c}}/\Msol)  > 10.48$ compared
with the profile of quiescent centrals (red solid line).
\emph{Right:} the projected radial profile around all centrals (black
dotted line) at $1 < z <3$ with stellar masses of
$\log(M_{\unit{c}}/\Msol)  > 10.48$ compared with the profile of
star-forming centrals (blue solid line). In each panel, the number in
parentheses gives the number of centrals in each subsample. The number
of satellites around quiescent centrals is higher on average than that
around star-forming centrals at $1.4\sigma$.}
\label{fig:sfqcompareunmatched}
\end{figure*}

\begin{figure*}[h]
\epsscale{1.0}
\plotone{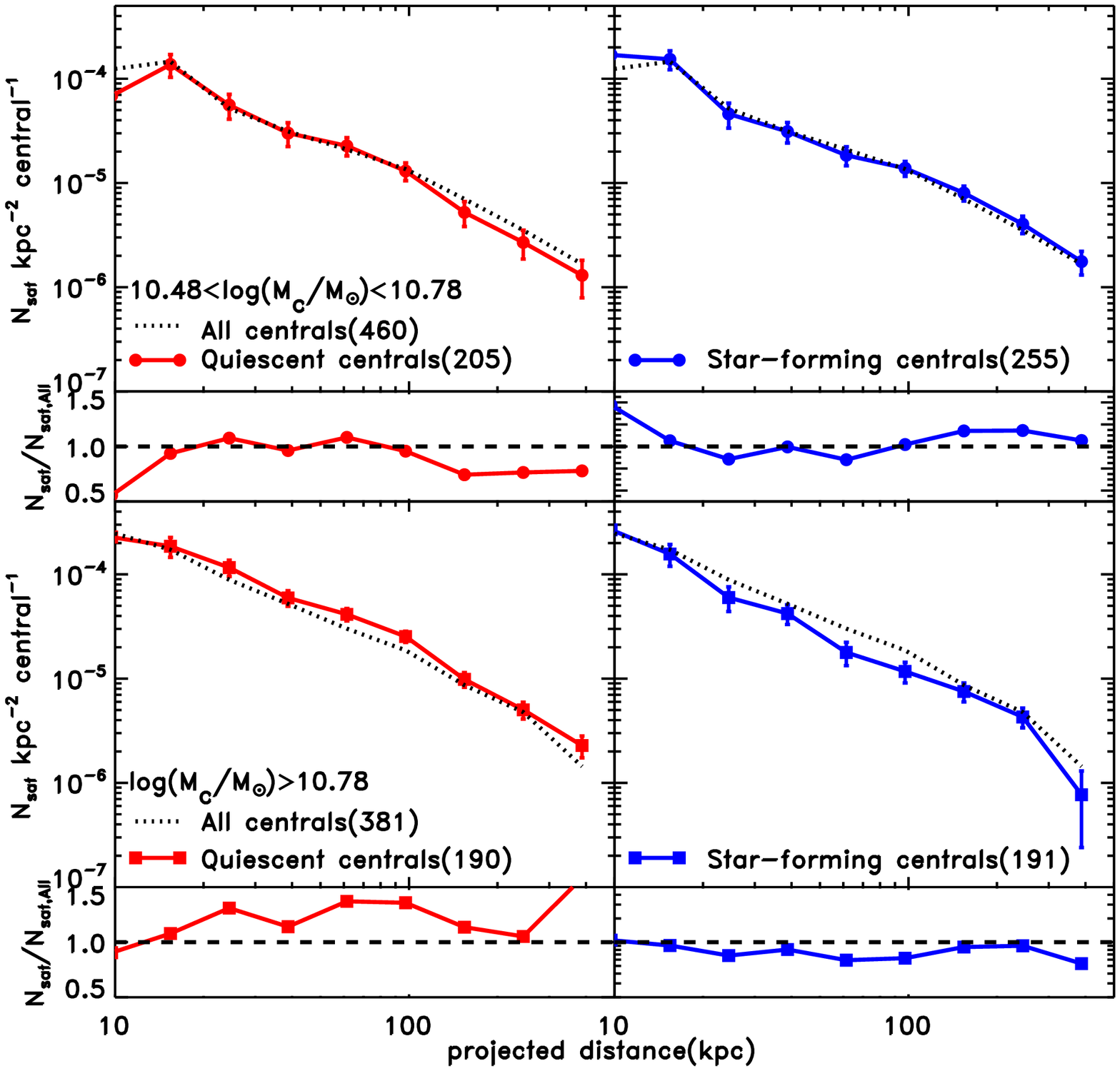}

\caption{ The dependence of the number density of satellites on
star-formation activity  and stellar mass. \emph{Top:} the projected
radial profile around all centrals (black dotted line) at $1 < z < 3$
with stellar masses of $10.48 < \log(M_{\unit{c}}/\Msol)  < 10.78$
compared with quiescent centrals (red solid line with circles, left
panel) and star-forming centrals (blue solid line with circles, right
panel).  \emph{Bottom:} the projected radial profile around all
centrals (black dotted line) at $1 < z < 3$ with stellar masses of
$\log(M_{\unit{c}}/\Msol)  > 10.78$ compared with quiescent centrals
(red solid line with boxes, left panel) and star-forming galaxies
(blue solid line with boxes, right panel). The number in parentheses
is the number of centrals in each subsample. Below each
$N_{\mathrm{sat}}$ plot the ratio of $N_{\mathrm{sat}}$ around
quiescent (star-forming) centrals to the $N_{\mathrm{sat}}$ around all
centrals is shown to illustrate the difference  between
$N_{\mathrm{sat}}$ around quiescent and star-forming centrals.  The
uncertainty on the ratio is derived assuming that the uncertainties on
the number of quiescent galaxies and star-forming galaxies both are
given by Poisson statistics, and they are uncorrelated. However, the
derived error bars of the ratios are very small compared to the size
of the data points. For the centrals with stellar masses of
$\log(M_{\mathrm{c}}/\msol)  > 10.78$, the quiescent central galaxies
have $\sim$ 2$\times$ the number of satellites compared to
star-forming centrals, and this difference is significant at
3.1$\sigma$.}
\label{fig:sfqmasscompareunmatched}
\end{figure*}

\subsection{Dependence on Star Formation Activity of Central
  Galaxy}

We investigate how the satellite distribution depends on the
star-formation activity of the central galaxies by dividing our sample
of central galaxies into subsamples that are star-forming and
quiescent (where these labels correspond to galaxies with high and low
sSFRs) using their rest-frame $U-V$ and $V-J$ colors as illustrated in
Figure~\ref{fig:uvjdiagram} and discussed in \S~\ref{sec:samplesel}.
We then recompute the satellite distribution for each subsample.

Figure~\ref{fig:sfqcompareunmatched} shows the projected radial
distribution of satellites around the star-forming and quiescent
centrals with $\log(M_{\unit{c}}/\Msol) > 10.48$. We find that
quiescent centrals host more satellites than their star-forming
counterparts.  Using our Monte Carlo simulation the likelihood we
would obtain this result by chance has a probabilty of $p_\mathrm{MC}=
0.081$, corresponding to 1.4$\sigma$ significance. 

\par We further investigate how the number density of satellites
depends on both star-formation activity and stellar mass of the
central.   Figure~\ref{fig:sfqmasscompareunmatched} shows for centrals
with moderate stellar mass  $10.48 < \log(M_{\mathrm{c}}/\msol)  <
10.78$. There is no significant evidence that the number density
depends on star-formation activity (with a $p$-value
$p_{\mathrm{MC}}=0.478$):  both quiescent and star-forming moderate
mass centrals have the same number of satellites.    In contrast, all
the difference in the number density of satellites occurs for centrals
at the high stellar-mass end.  For the high mass centrals,
$\log(M_{\mathrm{c}}/\msol)  > 10.78$, the quiescent central galaxies
have a significant excess of satellites compared to the star-forming
centrals, with a $p$-value of $p_{\mathrm{MC}} = 0.002$ (significant
at about $\simeq 3.1\sigma$ ). We discuss the implications of these
results in \S~\ref{sec:discussion}.

\begin{figure*}
\epsscale{1.0}
\plottwo{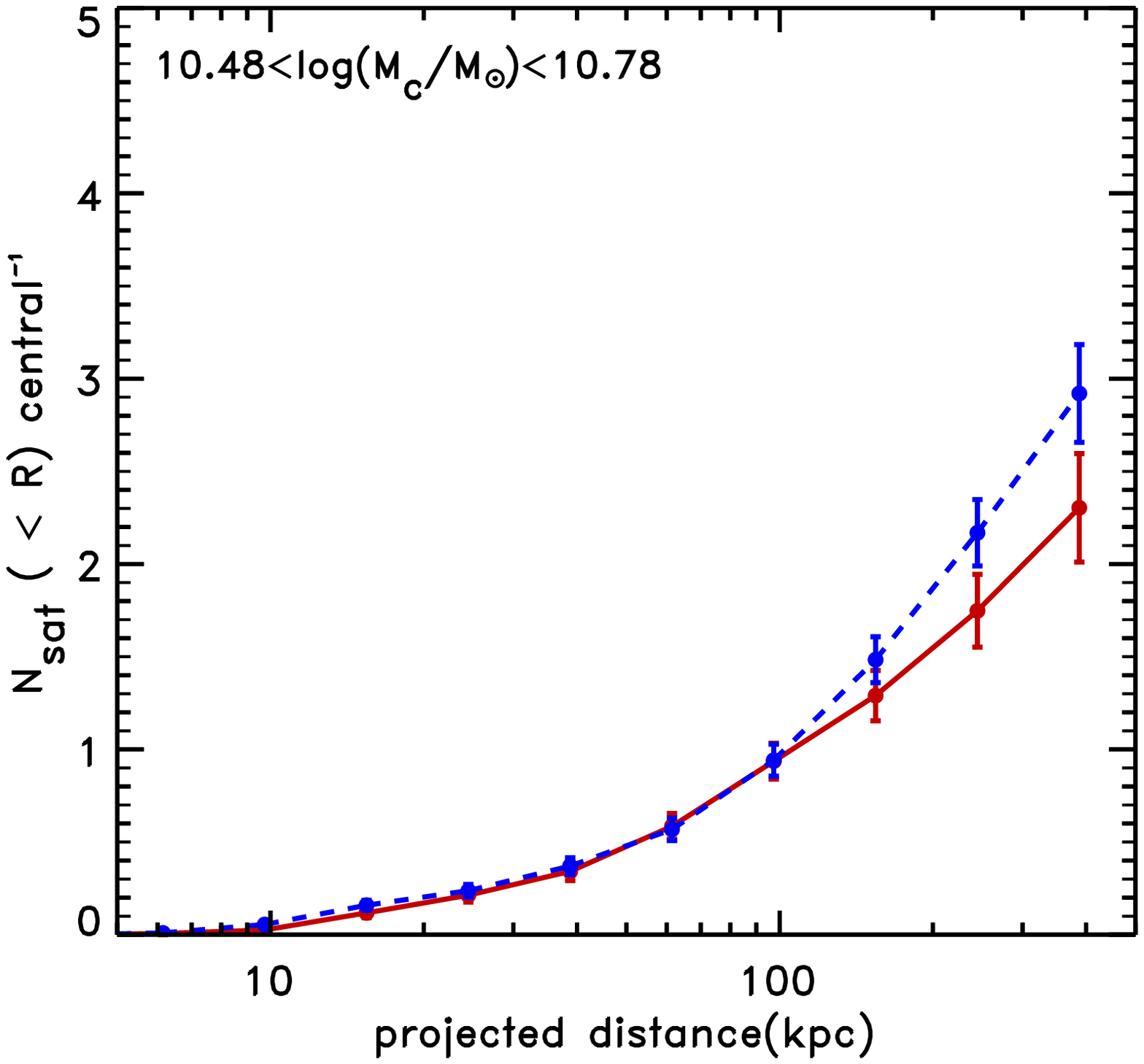}{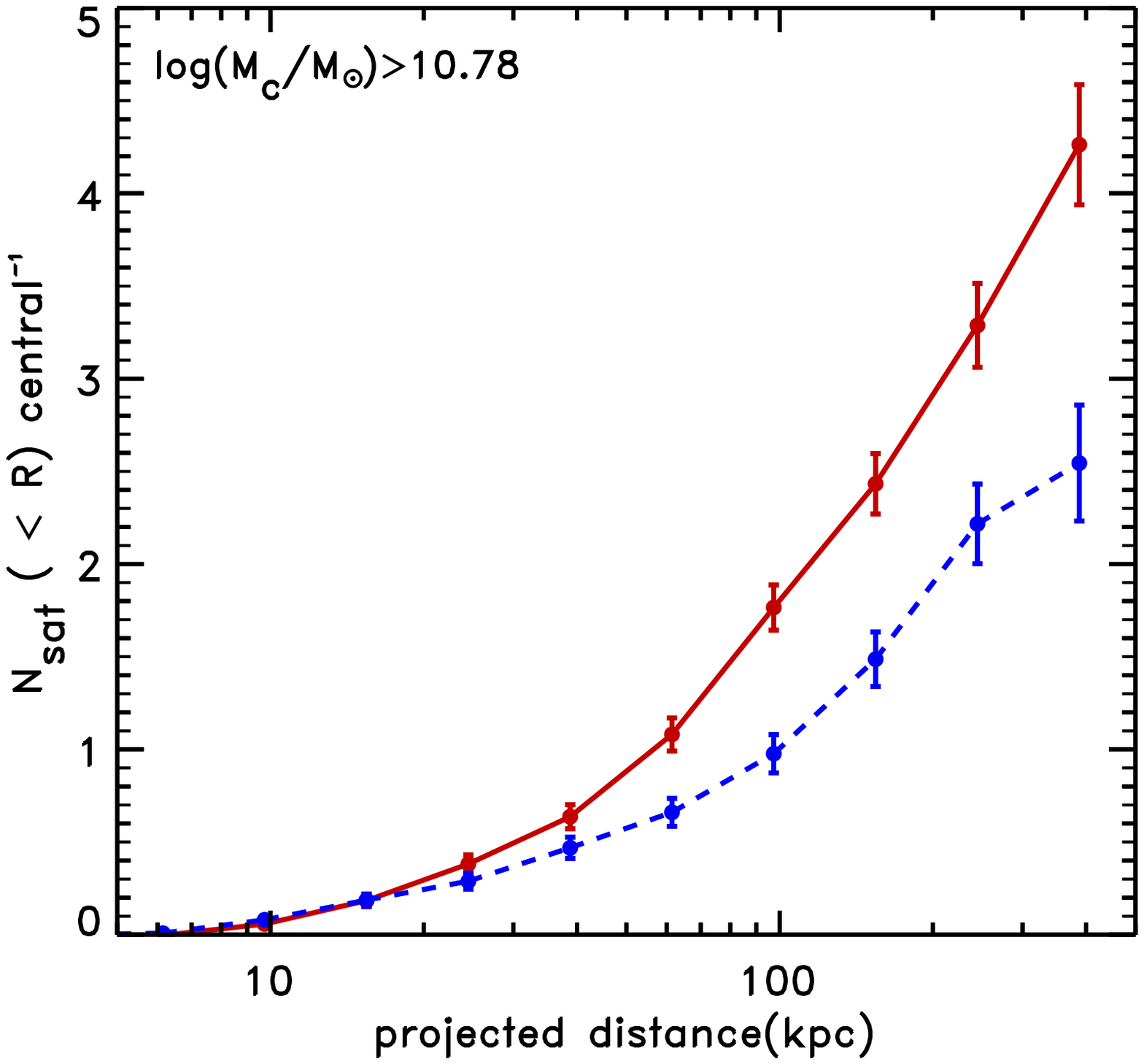}
%
\caption{Cumulative number of satellites as a function of projected
distance for quiescent central galaxies (red solid lines) and
star-forming central galaxies (blue dashed line). \emph{Left:} those
centrals with $10.48 < \log(M_{\unit{c}}/\msol)  < 10.78 $.
\emph{Right:} the cumulative number of satellites for centrals with
stellar mass of $\log(M_{\unit{c}}/\msol)  > 10.78$.   At $1 < z < 3$
centrals with these stellar masses have between 2 and 4
satellites within 400 kpc depending on mass and
star-formation activity.}  
\label{fig:cumudistunmatched}
\end{figure*}

\subsection{Cumulative Number Density of Satellite Galaxies}

We integrate the satellite number densities to measure the total (cumulative)
number of satellites within a projected distance of the centrals in
our samples down to our mass limit for the satellites, $\log
(M/\msol) > 9$.   Figure ~\ref{fig:cumudistunmatched} compares the cumulative
number density of satellites around quiescent centrals and
star-forming centrals with $10.48 < \log(M_{\unit{c}}/\msol)  <
10.78$ and $\log(M_{\unit{c}}/\msol)  > 10.78$.     On average the
intermediate mass centrals have $\approx$1 satellite more massive than
$\log(M_\mathrm{sat}/\msol) > 9$ within 100 kpc.  This is true for
both the star-forming and quiescent centrals.  The intermediate 
mass star-forming galaxies have an excess of satellites at larger
projected radii than the quiescent centrals, but this has $<$2$\sigma$ significance. 

Figure~\ref{fig:cumudistunmatched} also shows that the massive
quiescent centrals ($\log(M_\mathrm{c}/\msol) > 10.78$) have nearly
double the number of satellites more massive than $\log(M_\mathrm{sat}
/ \msol) > 9$ within 100--400 kpc compared to the massive
star-forming centrals. On average a massive star-forming central has
$\approx$ 2 such satellites, whereas a massive quiescent
central has $\approx$ 4.  These results are comparable with
the number of satellites found around massive centrals by
\citet{Tal2013}, who find that on average the total number of galaxies
with the mass ratio of  1:10 and within 400 $\unit{kpc}$ around the
massive centrals between $z=0.04$ to $z=1.6$ is 2 to 3.


\begin{figure*}
\epsscale{1.0}
\plottwo{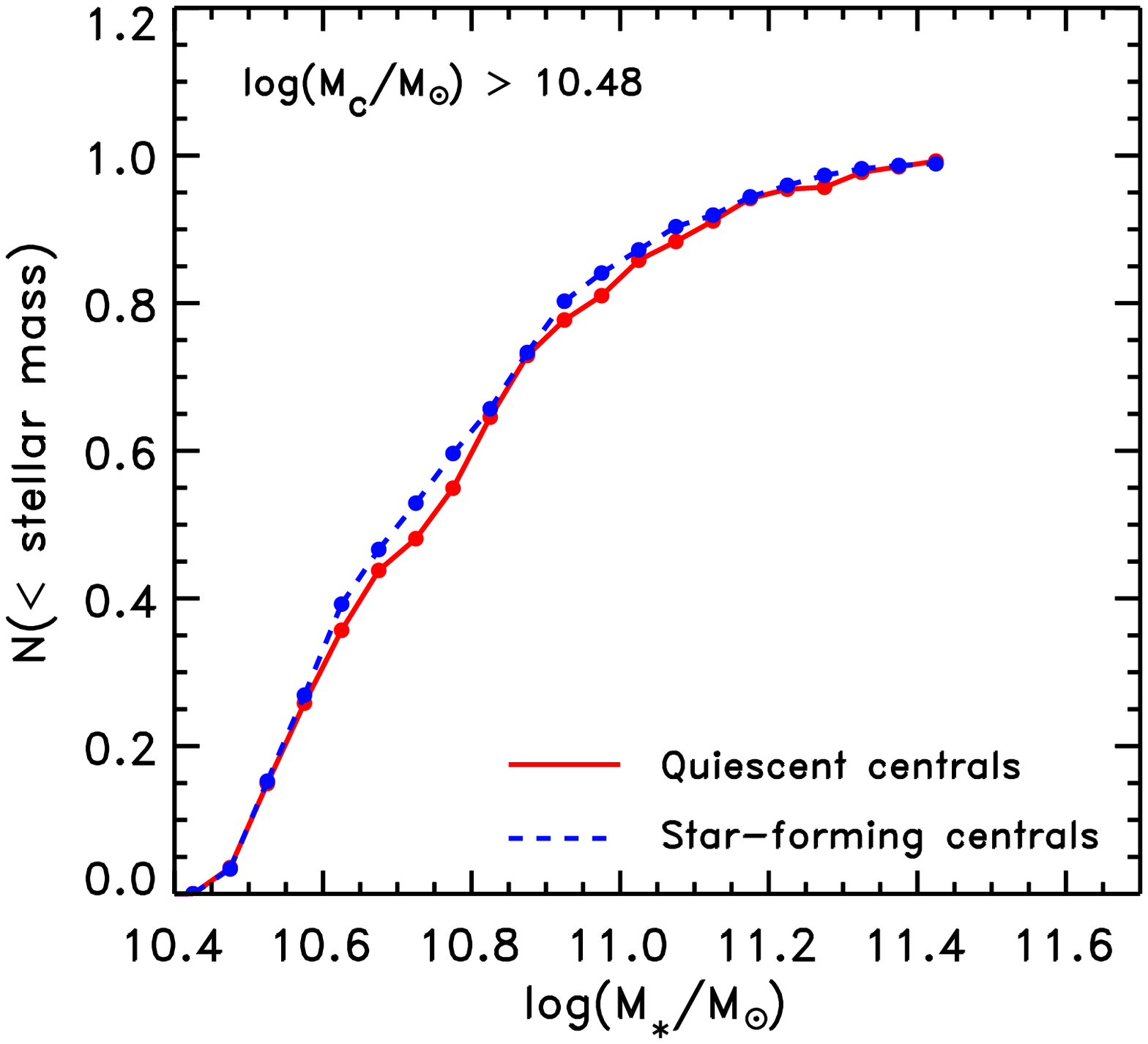}{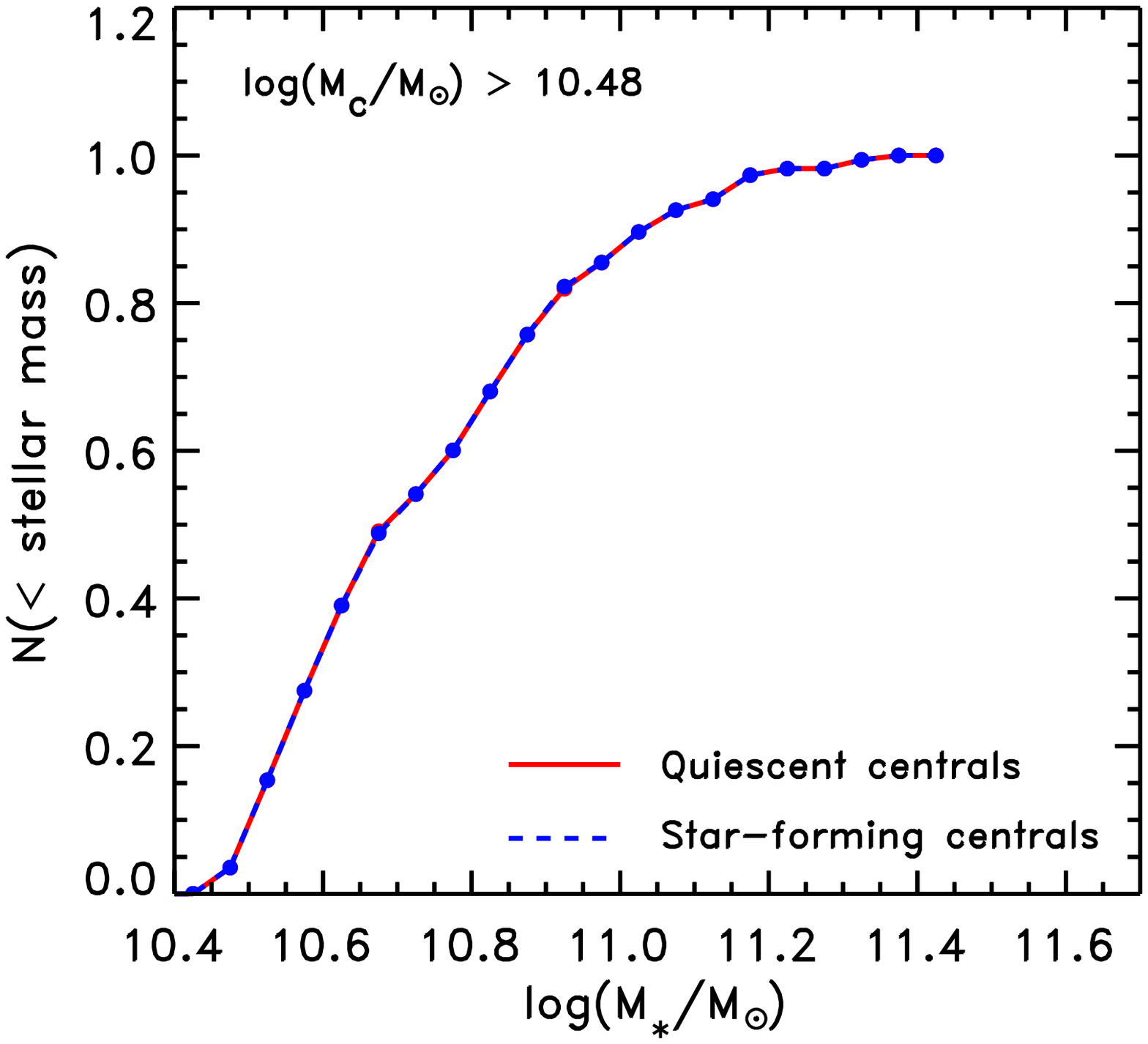}
\caption{Cumulative distribution in stellar mass for the quiescent
  centrals (red curves) and star-forming centrals (blue curves), for
  centrals $\log (M_\mathrm{c}/\msol) > 10.48$. \emph{Left:} the stellar
  mass distribution of the full samples in our \zfourge\
  data.   The quiescent centrals have a slightly higher median stellar
  mass, and this could be related to those galaxies having more
  satellites than the star-forming galaxies. \emph{Right:} the same 
  distribution after we have matched the stellar mass
  distributions.  This allows us to test if the quiescent centrals
  have more satellites even when they are matched in stellar mass to
  the star-forming centrals. }
\label{fig:cumumassunmatched}
\end{figure*}

\section{Discussion}
\label{sec:discussion}
\subsection{Why do Quiescent and Star-forming Centrals have Different Satellite Distributions?: the Effects of Stellar Mass and Redshift}

There is significant evidence that at high stellar mass ($\log(M_{\unit{c}}/\msol) > 10.78$) quiescent galaxies have more satellites than star-forming galaxies.
There are two interpretations. One is that there is something intrinsic to a galaxy
being quiescent that also causes the galaxy to have more satellites.
The second is that the higher number of satellites is related to
stellar mass.  If the quiescent centrals have higher stellar masses than
the star-forming galaxies---even though the stellar mass \textit{limit}
is the same---then they may also be expected to have more satellites.

We test the second of these possibilities by looking at the cumulative
stellar mass distribution of the quiescent and star-forming
centrals.    As shown in Figure~\ref{fig:cumumassunmatched}, indeed
the quiescent centrals have a slightly higher median stellar mass,
which is higher than the median for the star-forming centrals by 0.05
dex.   

Although this difference in stellar mass between the quiescent and
star-forming centrals is small, it could affect the number of
satellites. Therefore, we make a new sample of quiescent centrals from our real
sample. First, we divide the sample of centrals into narrow stellar
mass bins. In each stellar mass bin we randomly select equal numbers of quiescent and star-forming galaxies, therefore creating a new sample 
matched in stellar mass. The numbers of the quiescent and star-forming
centrals in the matched stellar mass samples are shown in Table 2. 
The right panel of Figure~\ref{fig:cumumassunmatched} shows that the stellar mass distributions of the matched samples agree very well.

\begin{figure*}
\epsscale{1.0}
\plotone{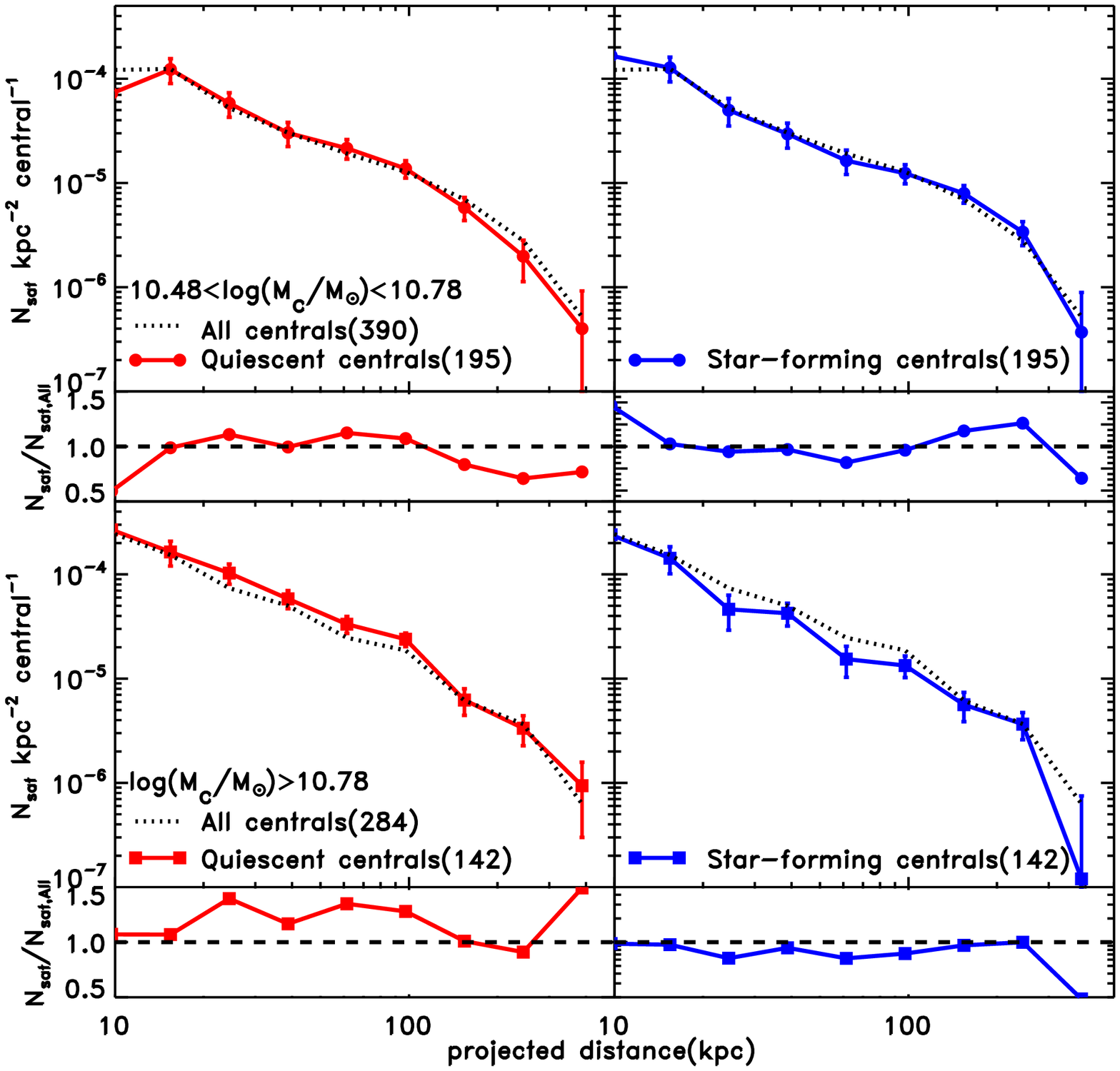}
%
\caption{Same as Figure ~\ref{fig:sfqmasscompareunmatched} but  for
the subsamples in which the cumulative stellar mass distribution of
quiescent and star-forming centrals have been matched (see
Figure~\ref{fig:cumumassunmatched}).  In the high mass subsample
$\log(M_{\unit{c}}/\Msol)  > 10.78$,  quiescent centrals have a higher
number density of satellites compared  to star-forming counterparts at
$2.7\sigma$ even when the stellar mass distributions of the quiescent
and star-forming galaxies are fixed.}
\label{fig:compmasstypefixedstellar}
\end{figure*}

After we match the mass distributions of quiescent and
star-forming centrals, we then recalculate the number density
profiles of satellites around the centrals for each subsample. 
Figure~\ref{fig:compmasstypefixedstellar} shows that the number densities of
satellites of moderate mass quiescent and star-forming centrals with
$10.48 < \log(M_{\unit{c}}/\msol) < 10.78$ are nearly identical with
no evidence for any difference.   Our Monte Carlo tests (\S~\ref{sec:errorestimate}) give
a 98.5\% likelihood ($p_{\mathrm{MC}}=0.985$) that the distributions are identical.

However, the excess of satellites around the massive quiescent
centrals with $\log(M_{\unit{c}}/\msol)> 10.78$ compared to the
massive star-forming centrals is still significant, where our Monte
Carlo tests give a likelihood that we would have obtained this result
by chance as 0.4\% (i.e., the difference is significant at $\simeq
2.7\sigma$ ($p_{\mathrm{MC}}=0.004$)). Therefore, while the offset in 
the stellar mass accounts for some of
the increase in the number of satellites around the massive centrals,
it is unable to account for all of it.  Even though the stellar masses are
matched, the massive quiescent centrals have more satellites than star-forming
centrals.

As another check, one could expect that quiescent and star-forming
galaxies may have different redshift distributions, i.e., if at fixed
stellar mass the star-forming galaxies  lie at higher redshift, then
this could possibly affect our results, as the number of satellites
(and dark matter halo mass) could build up with time.  For example,
\citet{Moster2013} show that at fixed stellar mass the halo mass of
massive galaxies increases with decreasing redshift.   

\begin{figure*}
\epsscale{1.0}
\plottwo{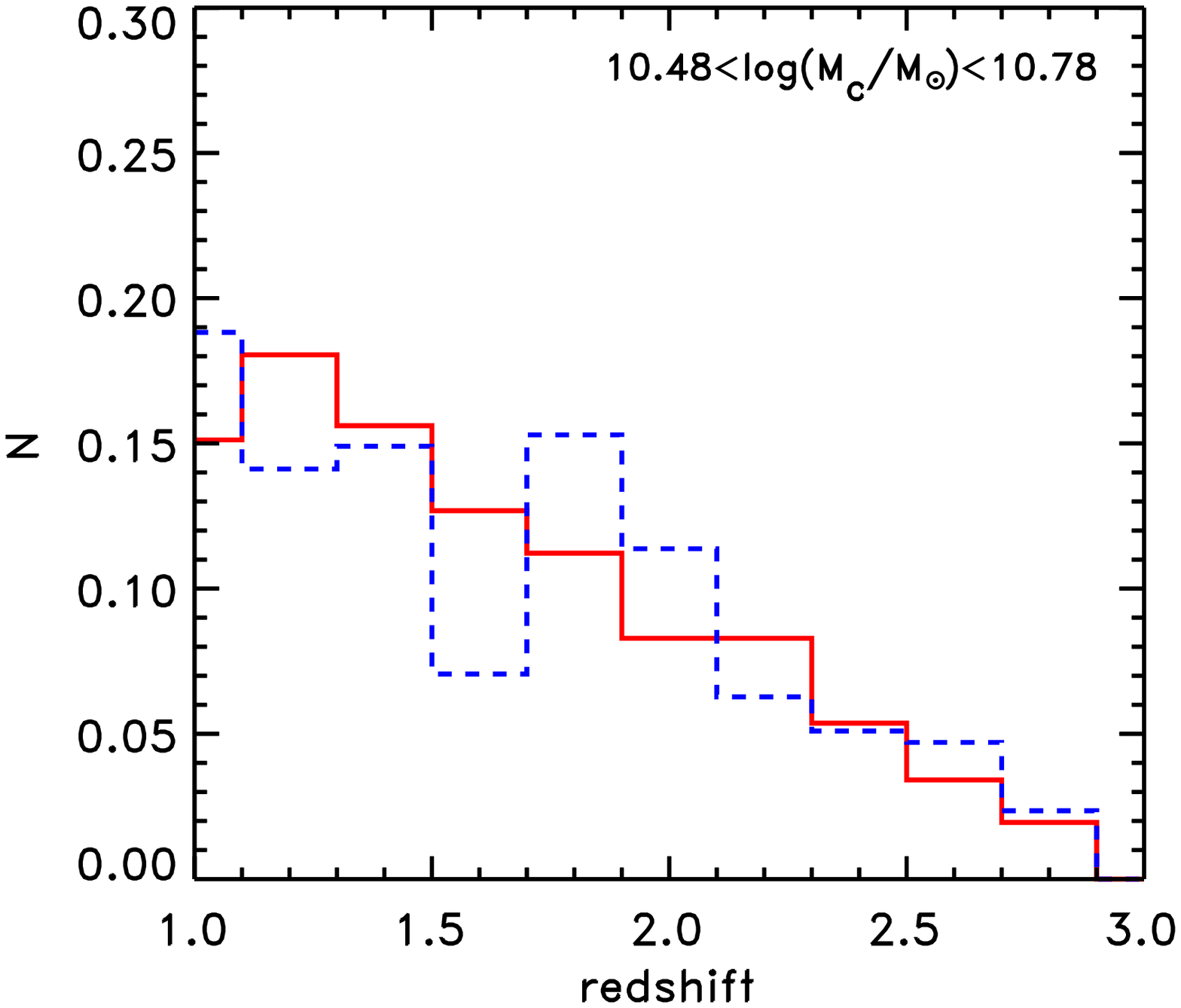}{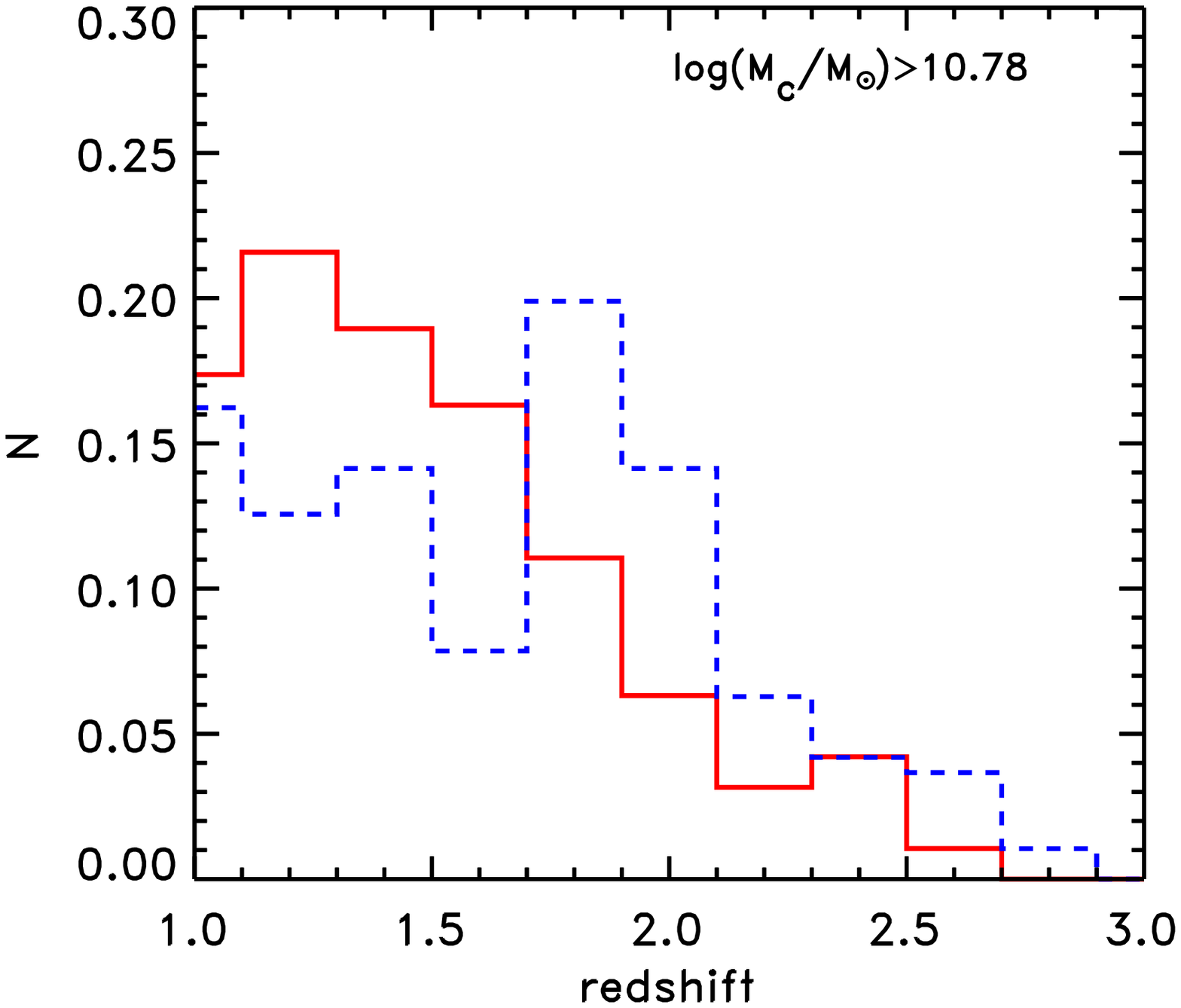}
\caption{\emph{Left:} The redshift distribution of quiescent (red
solid lines) and star-forming centrals (blue dashed lines) at $1<z<3$
for intermediate mass, $\log M/M_\odot = 10.48-10.78$. There are no
statistically significant difference between quiescent/star-forming
redshift distribution. \emph{Right:} The redshift distribution of high-mass centrals with $\log M_c / M_\odot > 10.78$.  The distribution of star-forming high-mass centrals is shifted toward higher redshift compared to the quiescent high-mass centrals.}
\label{fig:redshiftdist}
\end{figure*}

Figure~\ref{fig:redshiftdist} compares the redshift
distributions of quiescent and star-forming centrals for moderate mass
and high mass centrals. Using the WMW statistic we find no
statistically significant difference between the
quiescent/star-forming redshift distributions for moderate mass
centrals ($p_{\mathrm{WMW}} = 0.46$). The median redshift of quiescent
and star-forming centrals are 1.63 and 1.71, respectively. For
high-mass centrals the quiescent/star-forming redshift distribution is
significantly different ($p_{\mathrm{WMW}} = 0.002$). The median
redshift of quiescent and star-forming centrals are 1.57 and 1.79,
respectively. This difference is not a surprise, as the quiescent
fraction of galaxies at fixed mass increases with decreasing
redshift.

In order to test whether this difference in redshift
distributions affects our results, we match the redshift distributions
of high-mass quiescent and star-foming centrals and recalculate the
number density profiles of satellites for each subsample. We find that
the difference in the redshift distributions does not significantly
change our main result: there are still more satellites around massive
quiescent centrals compared to their star-forming counterparts.

As a final check, we recalculate the number density of satellites for
our samples of centrals, restricting the redshift range of centrals
to $1 < z < 2$.    The results are consistent with the satellite
distribution measured for the full $1< z < 3$ samples.  This also
implies that the number of satellites does not change very much over
this redshift range. \citet{Tal2013} find that the radial number
density of satellites has not evolved much over $z=0.04-1.6$. 
Therefore, our results show that the trend observed by Tal et al. extends 
to $z\sim 3$.

To summarize, there appears to be some 
physical connection between the quenching of star-formation and the 
presence of an increased number of satellites, at least for massive galaxies.    
One likely explanation is that the higher number of satellites corresponds to
larger dark-matter halo masses, and that at fixed stellar mass the
quiescent galaxies have higher halo mass.   Because we have no direct
measures of the halo masses of the galaxies in our sample, we test
this conclusion using semi-analytical models of galaxy formation in
the next section.
 
\subsection{Comparison to the Guo et al.\ Semi-Analytic Model: the Role of Halo Mass}
\label{sec:SAM}

To further explore the physical reasons that the massive
($\log(M_{\unit{c}}/\msol) > 10.78$) quiescent centrals have more
satellites than star-forming counterparts at $1 < z < 3$, we use
predictions from the semi-analytic model (SAM) of \citet{Guo2011}.
The Guo et al.\ SAM is derived using the Millennium-I simulation
\citep{Springel2001}. \citet{Henriques2012} provide mock ``lightcone''
catalogs from the Guo et al.\ models, and these lightcones include
galaxies at redshifts and to (low) stellar masses comparable to our
\ZFOURGE\ dataset.  

We select central galaxies from the mock catalogs using the same
redshift and stellar mass limits as for our \zfourge\ samples. We
further split the mock centrals by sSFR into quiescent
($\log(\mathrm{sSFR}/\mathrm{yr}^{-1}) < -10$) and star-forming
($\log(\mathrm{sSFR}/\mathrm{yr}^{-1}) > -10$) subsamples.  We use the
sSFRs for this classification because currently the
\citet{Henriques2012} light cones do not include rest-frame magnitudes
(e.g., we are unable to classify them using the $UVJ$ colors as done
for the \zfourge\ galaxies).   However, this makes little difference
as \citet{Papovich2012} show that at $z\sim 1.6$ the sSFR threshold of
$\log (\mathrm{sSFR}/\mathrm{yr}^{-1}) = -10$ effectively separates
galaxies classified as quiescent or star-forming by a $UVJ$-type
color-color selection. Therefore, the sSFR selection here is
equivalent to our $UVJ$ color-color selection above.  

\begin{figure*}
\epsscale{1.0}
\plotone{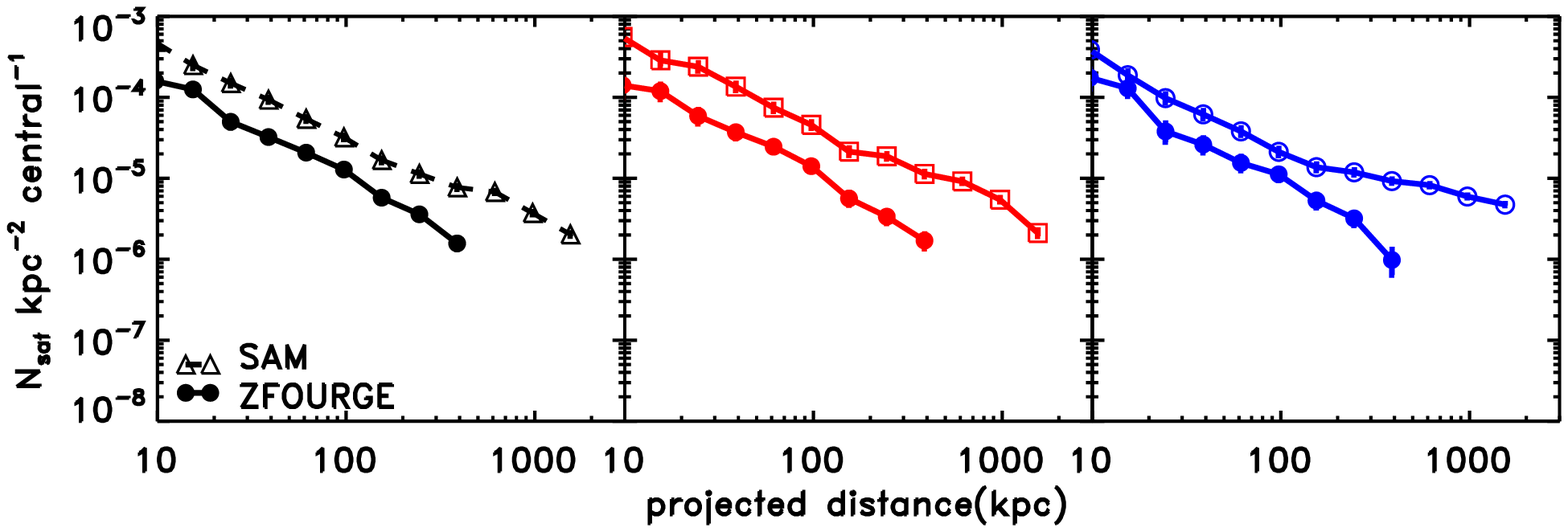}
%
\caption{The projected radial profile of satellites around each type
of central at $1 < z < 2$ with stellar masses of
$\log(M_{\unit{c}}/\msol)> 10.48$ for centrals selected from the
semi-analytic model (SAM; open circles) compared to the satellites
around the centrals in our \zfourge\ dataset (filled circles).  The
left panel shows the satellites for all centrals, the middle panel
shows the results for quiescent centrals, and the right panel shows
the results for star-forming centrals. In all cases the shape of the
satellite distribution is similar for the centrals in the SAM and
data, but the normalization is higher for the centrals in the SAM.}
\label{fig:compsimreal}
\end{figure*}

We identify centrals and measure the number density of satellites  at
$1 < z < 2$ in the SAM lightcone using the same methods as  applied to
the data. We restrict ourselves to comparisons between the SAM and our
data to  $1 < z < 2$ and $\log(M/\msol) > 9.33$ because this is the
adopted stellar mass-completeness limit for red satellites at $z = 2$
in the \zfourge\ data \citep{Tomczak2014}.  We note that the SAM is
also complete to this mass limit.  We then measure the projected
radial number density of satellites around centrals in the SAM and in
the data using this mass limit and redshift range. 

Figure~\ref{fig:compsimreal} compares the satellite number density
profiles in the SAM for the different central samples in our
\zfourge\ data.  The shape of the distributions is similar between the
SAM and the data, but the SAM has $\sim$3$\times$ the satellites than
the data at nearly all projected radii.  For our comparison here, we
are interested in the relative difference between the quiescent and
star-forming centrals in the data and the simulation, so this offset
is less important.  The reason for this offset is an interesting
problem (this is similar to the well-known ``missing satellite
problem'' \citep{Bullock2010}), and may indicate a mistreatment of
important physics in the models.  For example, the stellar mass
functions in the SAM show a higher number density of lower mass
galaxies at $1.3 < z < 3.0$ compared to observations (see Guo et
al.\ 2011, their Figure 23), and it may be expected that such a
disagreement would carry over to the satellite population.  However,
it is interesting that \cite{Wang2012, Wang2014} find a good agreement in the
abundance of satellites in the low-redshift SDSS data and in the
\citet{Guo2011} SAM, while we find such differences at higher redshift.
It is plausible that the higher abundance of low-mass galaxies
($\log(M/\msol) < 9.5$) at $z>1$ in the Guo et al. model has been
resolved by Henriques et al. (2013, see discussion therein).
Henriques et al.\ comment that in the Guo et al. model low-mass
galaxies form too early and are thus overabundant at high
redshift. \citet{Henriques2013} introduced modifications to the gas
reincorporation time-scales and produced an evolving galaxy population
which fits observed abundances as a function of stellar mass and
luminosity functions up to $z=3$. This may help resolve the abundance
of satellites as well.  

We fit the projected NFW profiles to the satellite distributions of
the centrals in the SAM and the data  (using the restricted redshift
range, $1 < z < 2$, and higher stellar mass-completeness  limit for
the satellites).    The fit to the projected NFW profile for the
satellite distribution in the SAM gives $r_s = 76.93 \pm 6.42
\unit{kpc}$. This is in reasonable agreement with the one we
measure for the comparable sample in the \zfourge\ data, where the fit
gives $r_s = 66.73 \pm 10.69 \unit{kpc}$ (now restricted to
the redshift range $1 < z < 2$, which accounts for the differences
between $r_s$ derived here an that for the full redshift range in \S~3.2).


\begin{figure*}
\epsscale{1.0}
\plotone{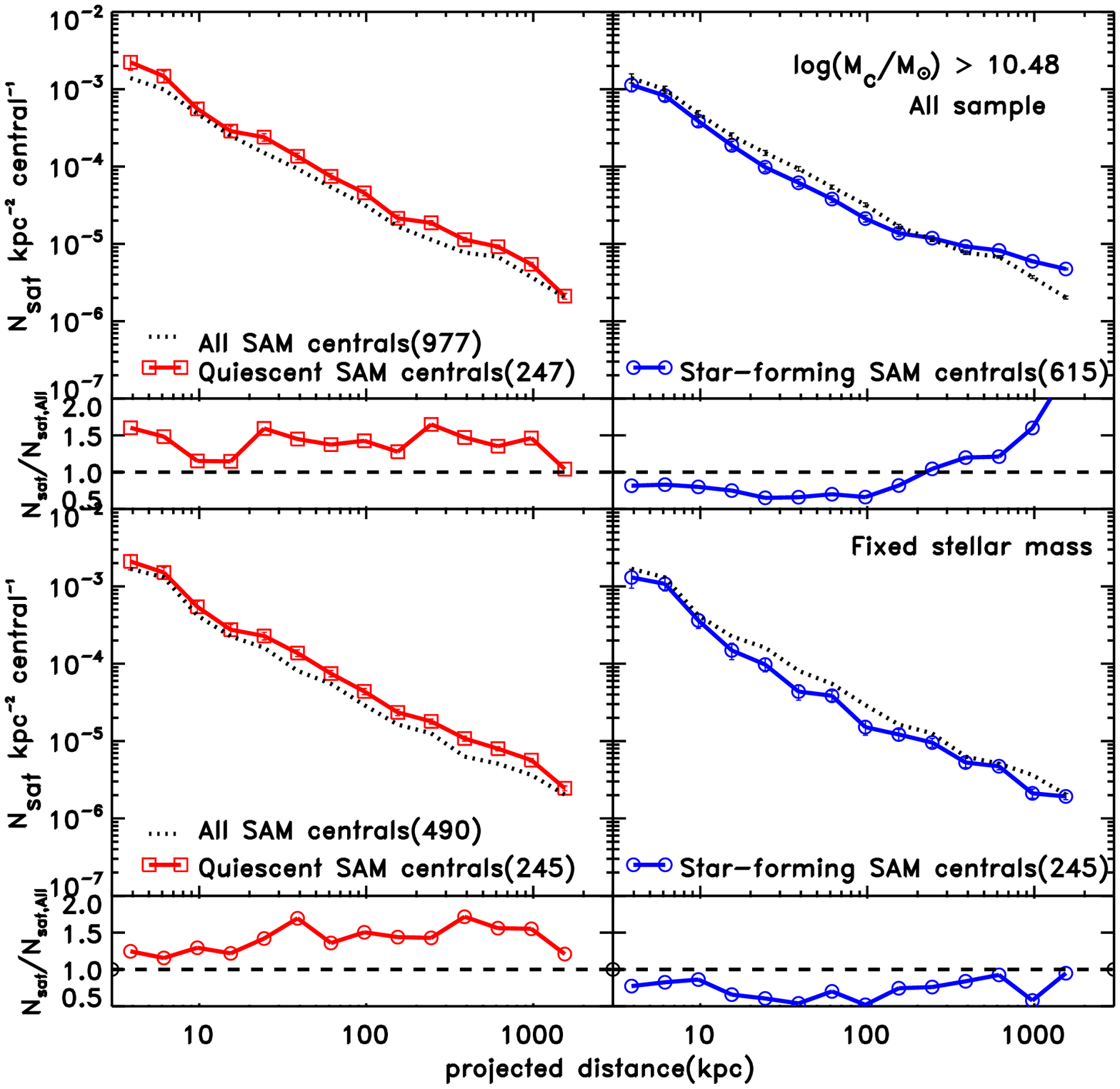}
%
\caption{\emph{Top:} the projected radial profile of satellites around
all centrals in the SAM (black dotted line), quiescent centrals (red
solid line with boxes in the left panel), and star-forming centrals
x(blue solid line with filled circles in the right panel) at $1 < z <
2$ with stellar masses of  $\log(M_{\unit{c}}/\msol) >
10.48$. \emph{Bottom:} same plot as the top panels but where we have
matched the stellar mass distributions of the quiescent and
star-forming centrals. In each panel, the number in parentheses
indicates the number of centrals in each subsample. There is a
significant  excess of satellites around quiescent central even when
we match the stellar mass distributions of the centrals.}
\label{fig:nsat2cases}
\end{figure*}

\begin{figure*}
\epsscale{1.0}
\plottwo{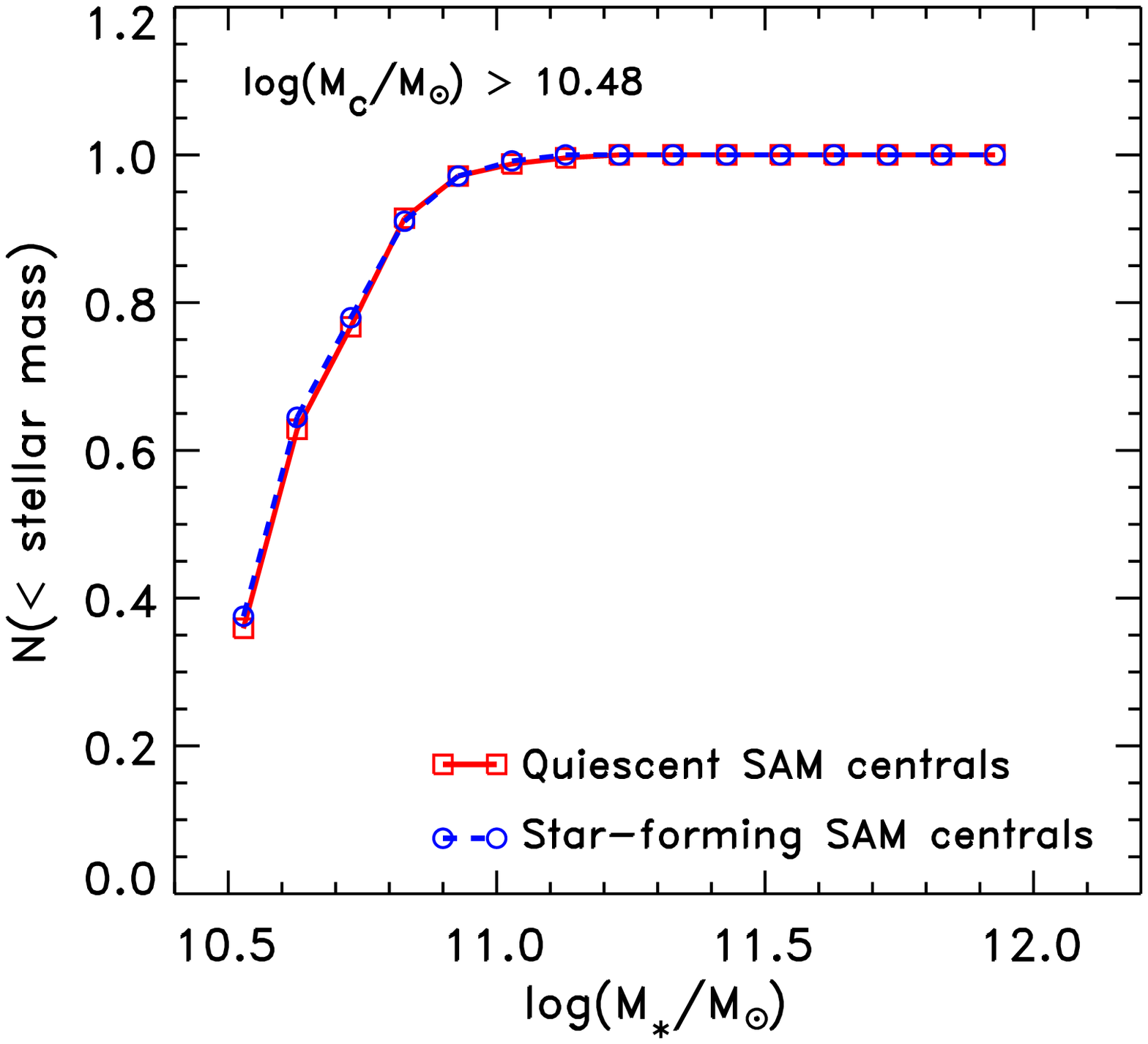}{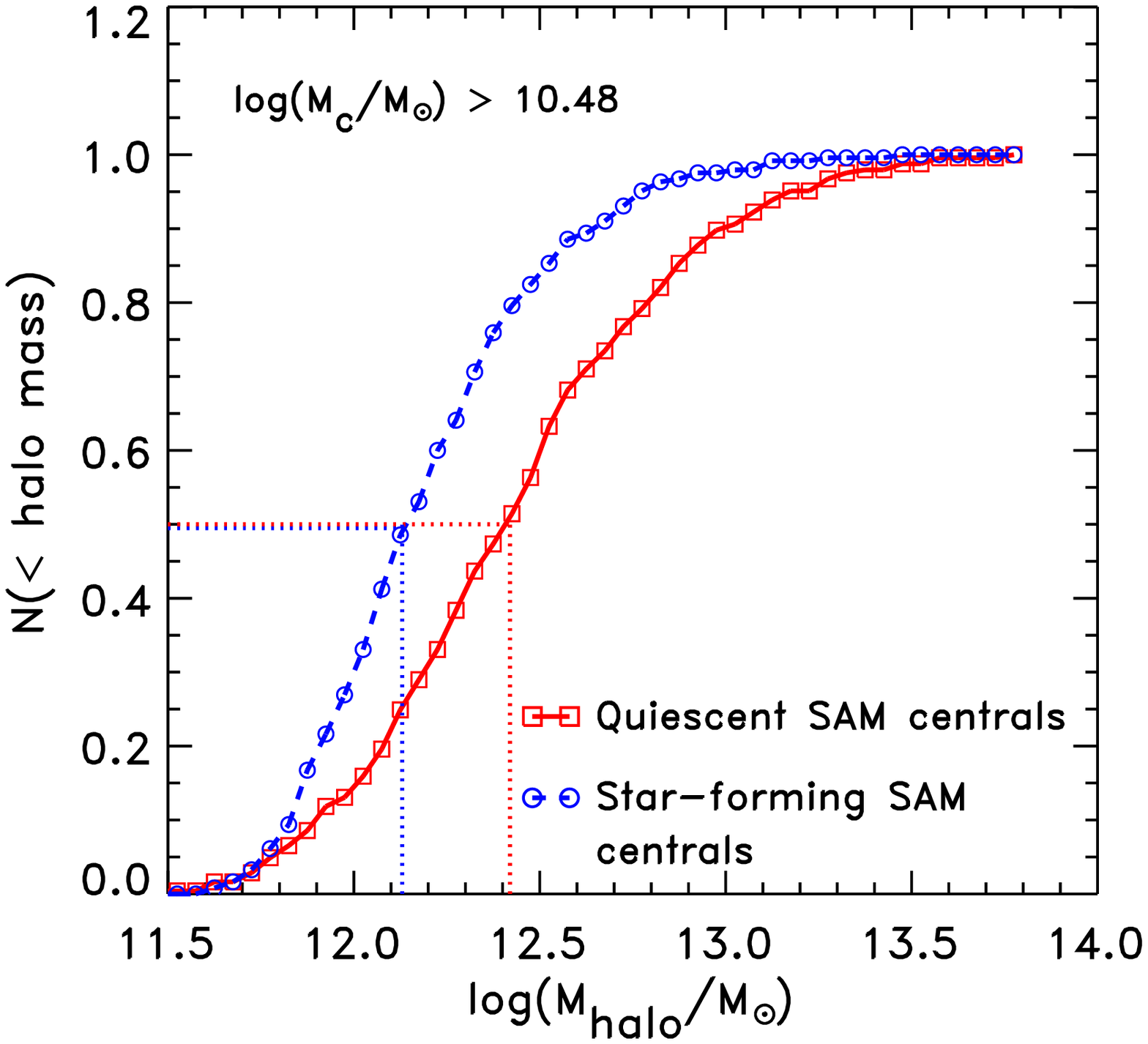}
%
%
\caption{ \emph{Left:} the cumulative stellar mass distribution for all
quiescent centrals and star-forming centrals in the SAM after
matching their stellar mass distributions. \emph{Right:} the
cumulative distributions of the halo masses of quiescent (red solid
curve) and star-forming (blue dashed curve) centrals after matching
their stellar mass distributions (as shown in the left panel).
Even when the centrals are matched in stellar mass, the quiescent
galaxies have higher halo masses.  The dotted lines show that the
median halo mass is higher by $\approx$0.3 dex.}
\label{fig:matchedcumsim}
\end{figure*}

\par To compare with the data, we investigate how the number density
of satellites around centrals in the SAM depends on stellar mass and
star-formation activity.  Figure~\ref{fig:nsat2cases} shows the number
density of satellites for all the SAM centrals with
$\log(M_{\unit{c}}/\msol) > 10.48$, and for the quiescent and
star-forming centrals separately.  As with the data, there is an
excess of satellites around quiescent galaxies and most of the signal
comes from the most massive centrals, with
$\log(M_{\unit{c}}/\msol)>10.78$: the $p$-values are
$p_{\mathrm{MC}}=0.032$  ($\simeq 1.9\sigma$) and
$p_{\mathrm{MC}}=0.001$ ($\simeq 3.7\sigma$) for the centrals with
$10.48 < \log(M_{\unit{c}}/\msol) < 10.78$ and
$\log(M_{\unit{c}}/\msol) > 10.78$, respectively. As with the data, we
investigate if the excess of the satellites  around the quiescent
galaxies is a result of a higher average stellar mass.    We therefore
match the stellar mass distributions between the star-forming and
quiescent centrals in the SAM (see
\S~5.1). Figure~\ref{fig:matchedcumsim} shows the cumulative stellar
mass distribution of the SAM centrals after the stellar mass
distributions are matched.    Figure~\ref{fig:nsat2cases} shows that
quiescent centrals in the SAM have a higher number density of
satellites compared to the star-forming centrals, even after the
stellar mass distributions have been matched. The $p$-values are
$p_{\mathrm{MC}}=0.050$  ($\simeq 1.7\sigma$) and
$p_{\mathrm{MC}}=0.016$ ($\simeq 2.1\sigma$) for the centrals with
$10.48 < \log(M_{\unit{c}}/\msol) < 10.78$ and
$\log(M_{\unit{c}}/\msol) > 10.78$, respectively.

\begin{figure*}
\epsscale{1.0}
\plotone{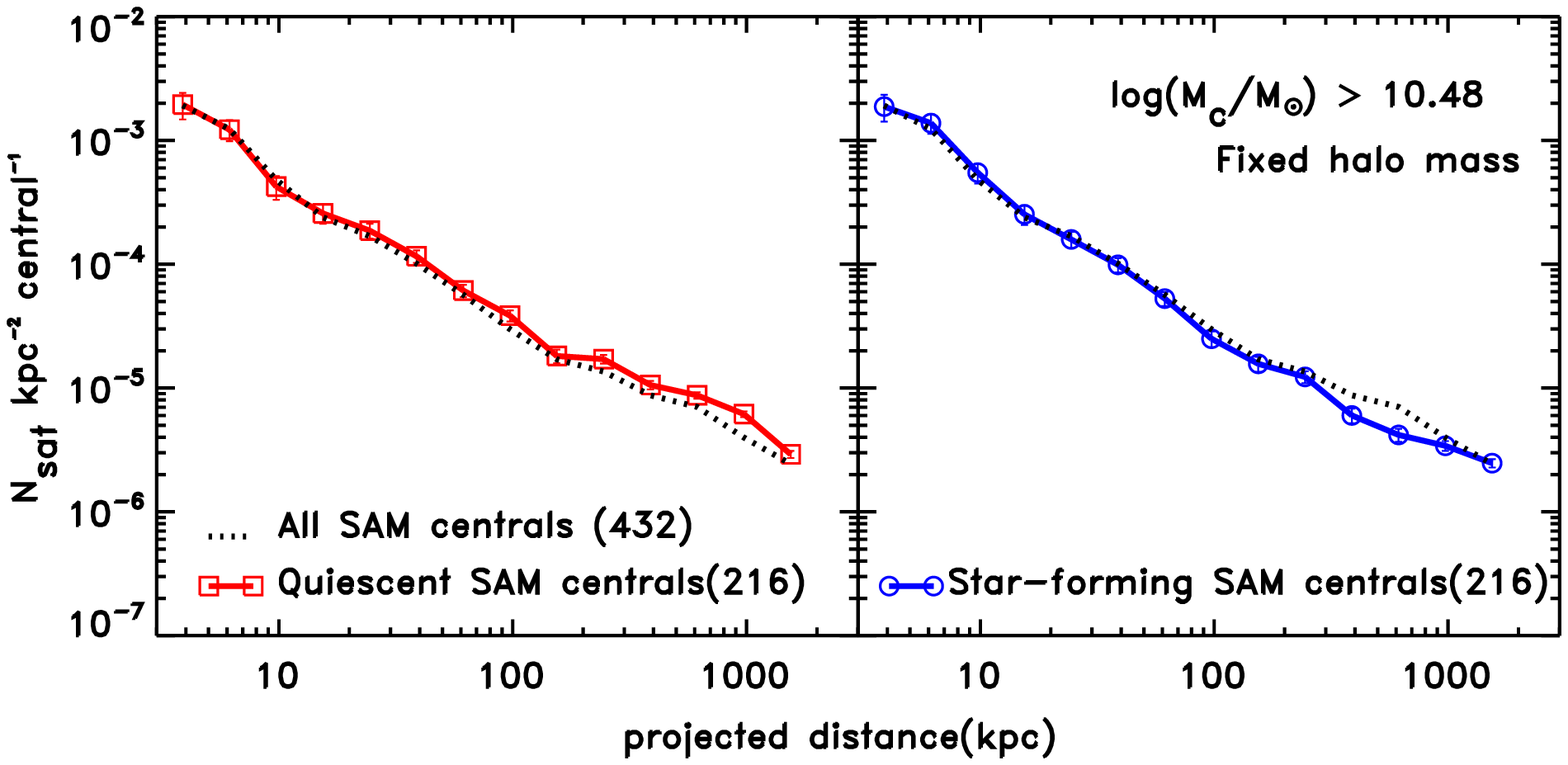}
%
\caption{Same as Figure ~\ref{fig:nsat2cases} but for the subsamples
in which the cumulative halo mass density of quiescent and
star-forming centrals in the SAM at $1 < z < 2$ with stellar masses of
$\log(M_{\unit{c}}/\msol) > 10.48$ have been matched.  As indicated in
the plot, fixing the halo mass makes the number density of satellites
at $10 < r/\mathrm{kpc} < 100$ equal between the star-forming and
quiescent centrals.   Therefore the excess number density of
satellites is an indication that the latter systems have higher halo
mass. }
\label{fig:nsat4cases3}
\end{figure*}

\par The difference in number density of satellites between quiescent
and star-forming centrals in the SAM is similar to that observed in
our centrals in the \ZFOURGE\ data. The massive quiescent centrals
($\log(M/\msol)>10.78$) have about twice the number of satellites
relative to massive star-forming centrals.  This is still the case in
a stellar mass-matched sample.  However, we do note that the
difference in satellite content can still be found in the SAM at
intermediate masses ($10.48 < \log(M_{\unit{c}}/\msol) < 10.78$),
whereas we do not see a significant  difference in the data.

It is a concern that our sample of centrals could have some contamination
from galaxies misclassified as centrals that are themselves satellites of
other centrals.   Since this contamination is expected to be worse at
intermediate masses than at higher masses, this may affect our results
on the mass-dependence of satellite content.  In particular, it could
contribute to the lack of observed difference in satellite content
between intermediate-mass quiescent and star-forming galaxies.  One way
to circumvent the problem of satellite contamination would be to apply
isolation criteria for centrals, however this would
reduce our sample size\footnote{We also note that isolation criteria
can introduce a bias against overdense regions, as discussed by
\citet{Wang2012} and \citet{Wang2014}. This bias is stronger when the
signal is weak, i.e. at large distances and for low-mass centrals.}.

Therefore, to test the magnitude of the effect of satellite
contamination in our central samples, we use the mock catalog,  in
which the true central and satellites are known.   We find that $\sim
72\% $ of all galaxies at intermediate mass ($10.48 <
\log(M_{\unit{c}}/\msol) < 10.78$) are centrals, while
($\log(M_{\unit{c}}/\msol) >10.78$) $\sim78\%$ of all galaxies in the
high-mass sample are centrals, where the rest are misclassified
satellites.  We excluded these ``false centrals'' and recalculated the
number density of satellites, again divided into subsamples of
quiescent and star-forming centrals.   For intermediate-mass centrals,
the number of satellites around both quiescent and star-forming
galaxies decrease by $\sim30\%$.  Because this affects both subsamples
nearly equally it does not change the relative difference between
them, and this should have no impact on our conclusions.    For high
mass centrals,  the number density of satellites declines by
$\sim40\%$ for star-forming centrals,  but only by $\sim10\%$ for
quiescent centrals.  Therefore, while we make no correction for
``false centrals'' in our analysis, we note that correcting for this
affect would \textit{reduce} the number density of satellites around
the high-mass star-forming centrals relative to the high-mass
quiescent galaxies, enhancing the significance of difference between
these subsamples measured in the data.

Therefore, we use the mock catalog from the SAM to investigate the
underlying reason for the excess satellites around massive quiescent
galaxies.  The main reason appears to be that in the SAM the quiescent
centrals have higher halo masses compared to the star-forming centrals
at fixed stellar mass.    Figure~\ref{fig:matchedcumsim} shows that
after we have matched the stellar mass distributions of the centrals
in the SAM, the quiescent centrals have a higher median halo mass by a
factor of $\approx 0.3$ dex (factor of order 2).     

To test if halo mass is the driving cause, we match the halo mass
distributions between the quiescent and star-forming centrals in the
SAM using the method to match the stellar mass distributions (see
\S~5.1). Figure~\ref{fig:nsat4cases3} shows the number density of
satellites around quiescent and star-forming centrals after matching
their halo mass distributions. The difference in the number density of
satellites has almost entirely disappeared.  Using our $p$-values from
Monte Carlo simulations (see \S~4.3), we derive a likelihood of
$p_{\mathrm{MC}} =0.176 $ ($\simeq 0.9\sigma)$ that there is a
difference between the satellite distributions for the centrals with
$\log(M_{\unit{c}}/\msol) > 10.48$. Therefore, in the SAM most of the
excess in the number density satellites around quiescent galaxies can
be attributed to those galaxies having higher halo masses compared to
star-forming centrals.

\par In $\Lambda$CDM, the dark matter halos grow through accretion and
mergers.  Clearly, this will involve the accretion and merging of
smaller halos that contain the satellite galaxies. Our analysis of the
SAM suggests that the observed difference in satellite content between
different types of galaxies is driven by differences in halo mass
\citep[see also][]{Cattaneo2006}, with the number of satellites
roughly proportional to the halo mass. Therefore, a plausible
interpretation of our results is that, at $\log(M_{\mathrm{c}}/\msol)
> 10.78$, quiescent centrals have a median halo mass that is about a
factor of 2 larger than comparable star-forming galaxies, and that
this difference becomes significantly smaller at lower masses.

\subsection{Comparison to Results at Lower Redshift}

Based on our analysis of the simulations in \S~\ref{sec:SAM}, we
interpret the excess of satellites around quiescent galaxies as
evidence that at fixed stellar mass quiescent centrals have more
massive dark matter halos than their star-forming counterparts.  Our
results are derived from galaxies from the three fields in ZFOURGE
(COSMOS, UDS, CDFS), which are well separated on the sky.  We see no
evidence for strong field-to-field variance (see
\S~\ref{sec:profilederive}), and therefore our results seem robust
against cosmic variance and/or systematics that vary between the
dataset in each field.

\par Our interpretation that quiescent galaxies have higher
dark-matter halos masses compared to star-forming galaxies agrees with
findings from analyses of galaxy clustering.  These studies also find
that quiescent galaxies have stronger clustering amplitudes, and
presumably higher dark mass halo masses, compared to their
star-forming counterparts \citep[e.g.,][]{Li2006,Hartley2013}.    It
is perhaps unsurprising that our results agree because our measurement
of the number distribution of satellite galaxies is similar to the
``one-halo'' term of the galaxy correlation function, but here we
measure this to much lower masses and because of our methodology we
are able to track the number of satellites and the mass contained
within them on average for each central galaxy.

Our results extend trends from the local Universe to higher redshifts.
For example, \citet{More2011} study the kinematics of satellite
galaxies from SDSS to infer the relation between the properties of
central galaxies and their halo masses. Similar to our findings, More
et al.\ conclude that central galaxies with the lower stellar masses
($\log(M/\unit{\Msol}) < 10.8$) have no significant difference in halo
mass regardless of being quiescent or star-forming. Moreover, More et
al.\ find that the more massive quiescent centrals have larger halo
masses compared to star-forming centrals even when the stellar mass is
fixed, again similar to our findings at high redshift.   More et al.\ find that
the difference between halo mass of quiescent and star-forming
centrals increases from 0.2 to 0.4 dex as the stellar mass of the
central increased from $\log (M/\unit{\Msol}) = 10.8$ to
$\log(M/\unit{\Msol})=11.1$. 

These similar results also have been found by other studies of low
redshift galaxies using the data from SDSS.\citet{Wang2012} study the
abundance of satellite galaxies in the stellar mass range $9.0
<\log(M/\unit{\Msol}) < 10.0$. They find that red centrals (what they
call ``primaries'') with the stellar masses of $\log(M/\unit{\Msol}) >
10.8$, have significantly more satellites than blue centrals of the
same stellar mass. For the centrals with stellar masses of
$\log(M/\unit{\Msol}) \sim 11.2$, red centrals have more satellites
about a factor of 2 relative to the star-forming counterparts. They
also compare the observation with the \citet{Guo2011} SAMs and find
that the red centrals have more satellites because they reside in more
massive halos. Recently, \cite{Phillips2014} study the satellites
around bright host galaxies with $\log(M/\unit{\Msol}) = 10.5$. The
distribution of velocity offset for satellites and their hosts show
that at fixed stellar mass the halo mass of passive host galaxies are
$\sim 45\% $ more massive than the those of star-forming galaxies.
 
These results are all in agreement with our findings. Therefore, it seems
as if there is little redshift evolution in the conclusion that
quiescent galaxies have higher halo masses than star-forming galaxies
at fixed stellar mass, at least for the more massive centrals. 

\subsection{Constraints on Models of Mass Quenching}

Qualitatively, it may not be surprising that quenched central galaxies
occupy more massive halos than star-forming galaxies at fixed stellar
mass, regardless of the particular quenching mechanism. Even after a
galaxy stops forming new stars, its halo will continue to grow at a
rate comparable to the past average
\citep[e.g.][]{Conroy2009,Moster2013}, meaning that the ratio of dark
matter mass to stellar mass will begin increasing relative to galaxies
that continue to form stars. This is consistent with the results we
derive from the \citet{Guo2011} SAM, where the quiescent galaxies have
higher median dark-matter masses compared to the star-forming
galaxies, even when we match the stellar mass distributions.

In this respect it is notable that we find a difference in the number
of satellites only in our high-mass sample ($\sim0.3$ dex), and no
significant difference at intermediate masses (upper limit $\sim0.1$
dex). Using the reasoning given above, this could be explained if high-mass
galaxies quench first, and thus their halos have had the most time to grow
relative to their stellar mass. Indeed, such mass--dependent quenching
has been clearly demonstrated by \citet[see their Figure 11]{Tomczak2014}.

\begin{figure*}
\epsscale{1.0}
\plotone{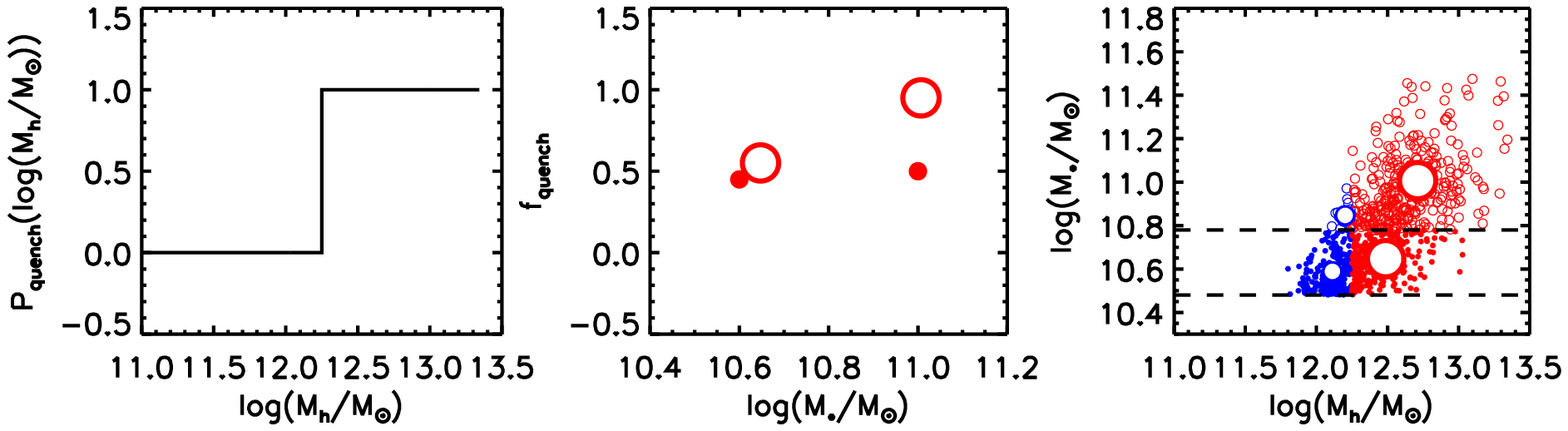}
\plotone{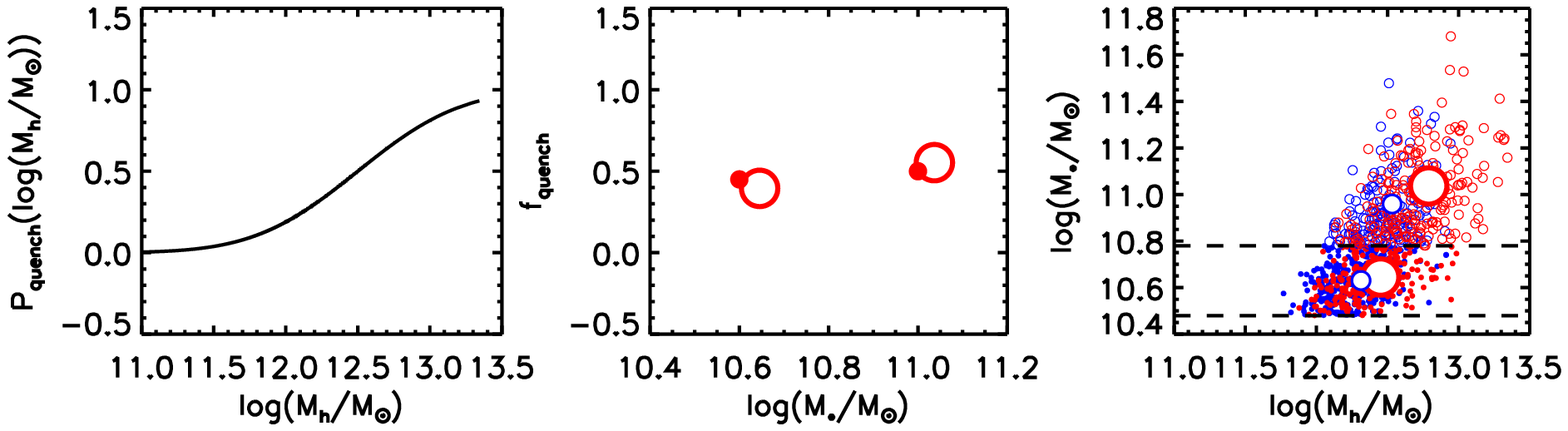}
\caption{A toy model that explores scenarios in which the
  quenching of star formation is related to halo mass, and which
  attempts to explain the large inferred difference in halo mass
  between star-forming and quiescent galaxies at high stellar masses,
  while maintaining a smaller difference at intermediate stellar
  masses. We use the redshift-dependent stellar-to-halo mass relation
  from \citet{Moster2013} to populate halos from an $N$-body
  simulation with galaxies, and add in additional scatter to the
  stellar masses. We then quench some subset of the galaxies based on
  their halo mass. \emph{Top:} A scenario where the quenching
  probability is a step function in halo mass, with a threshold of
  $\log(M_h/\msol)=12.25$, as shown on the left. The center panel
  shows that the predicted quenched fraction in our two stellar mass
  bins (large open circles) does not agree with the observations
  (small filled circles). The right panel shows the stellar and halo
  masses of the simulated galaxies, color-coded according to whether
  the galaxies have been quenched. The small and large open circles
  show the mean stellar and halo masses for star-forming galaxies and
  quiescent galaxies, respectively. The mean halo masses differ by 0.3
  dex in the intermediate mass bin and 0.5 dex in the high-mass bin,
  in contrast with the differences inferred from the data, which are
  $\sim0.08$ and $\sim0.3$ dex, respectively. \emph{Bottom:} same as
  the top panels for a scenario where the probability of each
  galaxy being quenched is modeled as a steplike function with a
  softened profile. The predicted quenched fractions, and the
  predicted differences in halo masses between star-forming and
  quiescent galaxies in our two mass bins, are in significantly better
  agreement with the data.}

\label{fig:toymodel}
\end{figure*} 

Our observations may have interesting implications for the
mechanisms that \emph{cause} quenching. It has long been recognized
that star formation in high-mass halos may be suppressed due to the
shock-heating, and the subsequent inefficient cooling, of infalling
gas \citep[e.g.][]{White1978,Birnboim2003,Keres2005,Dekel2006}. Here
we wish to interpret our results using a simple model in order
to test how halo mass and quenching are related. We first
use the redshift-dependent parametrization of the stellar-to-halo mass
relation from \citet{Moster2013} to populate halos at $1 < z < 3$ in
the Millennium simulation with galaxies, and add in 0.30 dex scatter
in stellar mass. We label the galaxies as star-forming if their halo
masses fall below a fixed threshold mass (a few times $10^{12} \msol$)
and quiescent if their halo masses fall above this threshold
(Figure~\ref{fig:toymodel}, top-left panel). We then calculate the
quenched fraction and the average halo mass of star-forming and
quiescent galaxies in our two mass bins (Figure~\ref{fig:toymodel},
top-center and top-right panels, respectively). This fixed halo-mass
threshold for quenching results in a quenched fraction that differs
significantly between the mass bins, which we do not observe. It also
predicts a mean halo mass of quiescent galaxies that is significantly
larger than the star-forming galaxies in both stellar mass bins, which
we also do not observe.

Because we cannot reproduce the observations for any single value of
quenching halo mass and scatter in stellar mass, rather than using the
model with a single halo mass quenching, we assign each galaxy a
probability for being quenched based on the halo mass.   We set the
probability to be a steplike function with a soft cutoff profile,
$P_{\mathrm{quench}}=0.5(1+\mathrm{erf}(\log(M_h/\msol)-\log
(M_{0.5}/\msol))/\sigma)$, where erf is the error function\footnote{
Our choice of functional form for $P_\mathrm{quench}$ here is
arbitrary except that it obeys our requirement that
$P_\mathrm{quench}$ increase with mass.  We expect other
parameterizations with a mass-dependent $P_\mathrm{quench}$ can
reproduce the data as well.}.  We adjust the parameter $M_{0.5}$,
which is defined such that $P_{\mathrm{quench}}(\log(M_{0.5}/\msol)) =
0.5$, and the parameter $\sigma$, as well as the scatter in stellar
mass to roughly reproduce the observed quenched fraction and the
difference in mean halo mass. We find that a $\log(M_{0.5}/\msol)$ of
$12.3-12.5$, a standard deviation of $0.7-0.9$, which corresponds to
$P_{\mathrm{quench}}=0$ for $\log(M_h/\msol) \sim 11-11.5$ and
$P_{\mathrm{quench}}=1$ for $\log(M_h/\msol) \sim 13.5$, and a scatter
in stellar mass of $0.15-0.2$ dex, is able to roughly reproduce the
observed quiescent fractions and the difference in the average halo
mass of centrals at both stellar mass bins (Figure~\ref{fig:toymodel},
bottom row). We find that the differences in mean halo mass of
quiescent and star-forming galaxies are $\sim 0.1$ dex and $\sim 0.2$
dex for intermediate and high stellar mass bins, respectively
(Figure~\ref{fig:toymodel}, bottom right panel), in better agreement
with the observations. The small scatter favored by the model in the
stellar-to-halo mass relation is due to the fact that we see
significant differences in $N_{\mathrm{sat}}$ over a relatively small
range in stellar mass: the mean mass in the intermediate-mass sample
is $\log(M_*/\msol) \sim 10.6$, while in the high mass sample it is
$\log(M_*/\msol) \sim 11.0$, and so a large amount of scatter would
wash out differences in halo mass over this relatively limited range
in stellar mass. We do note that the scatter in our modeling
represents intrinsic scatter in the stellar-to-halo mass relation
\emph{combined with} the random errors in our stellar mass estimates,
suggesting that the \emph{intrinsic} scatter must be small indeed.

Our model is simplistic, it is obviously possible to develop it
further. For instance, we note that adding a mass-dependent scatter in
the stellar masses --- where the scatter increases from 0.15 dex at
lower masses to 0.3 dex at the higher masses --- improves the
agreement between the toy model and the data. However, given the
uncertainties involved in the  current data, particularly with regard
to the incompleteness in satellite  detection at low masses and the
(limited) expected misclassification  between centrals and satellites
(see \S~2.2), our modeling results can only  be regarded as
indicative, and we do not push the modeling any further.

If our observation that intermediate mass quiescent and star-forming
centrals have the same $N_{\mathrm{sat}}$ is correct, then our toy
model strongly favors a scenario where there is no single quenching
halo mass threshold: even at relatively high halo masses
($\log(M/h^{-1}\msol) \sim 12$ based on the SAM), only about 50\% of
the galaxies are quiescent while the rest remain star-forming, and
galaxies have some likelihood of being quenched over a very wide range
in halo masses ($\log(M_h/\msol) \sim 11-13.5$). One remaining
question is that if halo mass quenching is an important mechanism,
\textit{why} do some galaxies remain star-forming while others quench?
Our result implies that halo mass can only be a contributing factor.
Other factors may include environmental processes (assembly bias and
environmental effects on the gas-accretion process), stochastic
processes such as mergers, and galaxy structure
\citep[e.g][]{Gao2005,Wechsler2006,Croton2007,Cooper2010,Papovich2012,Rudnick2012,Bassett2013,Lotz2013}.

This result may be expected on theoretical grounds, as some variation
in the quenching mass is expected due to variations in metallicity and
perhaps also due to the enhanced ability of cold flows to penetrate
halos at the higher redshifts in our sample
\citep{Dekel2006,Dekel2009}. A variation in halo mass is also expected
based on the results of \citet{Gabor2012}. In their model galaxy
quenching is based on the hot gas content of halos, which is
correlated with, but not directly tied to halo mass. Recently,
\citet{Lu2013} also show that galaxy models require a quenching
probability that increases with mass to explain the color-mass
distributions of galaxies in the CANDELS survey.

  
\par The mergers that grow massive quiescent galaxies are supposed to
be primarily dissipationless, and devoid of cold gas available for
star-formation \citep[e.g., so-called ``dry''
  mergers,][]{VanDokkum2010,Oser2010,Oser2012,Hopkins2010}. If this is
the case, it is expected that satellites around the quiescent centrals
in our sample, which will eventually merge with their central
galaxies, should be largely devoid of gas (or some process must cause
them to expel or consume their gas prior to merging with the central).
Therefore it may be expected that the satellites should show signs of
passive colors.  A discussion of the color and stellar mass
distributions of the satellites is beyond the scope of the present
work, but we will study these distributions in a future paper.

\section{Summary}

We have studied the statistical distribution of satellites around
star-forming and quiescent central galaxies at $1<z<3$ using imaging
from ZFOURGE  and CANDELS. The deep near-IR data allow us to select
satellites down to $\log(M/\msol)>9$ at $z<3$. The main conclusions of
this work are the following.

\begin{itemize}

\item  The projected radial number density of
satellites around centrals with stellar mass $\log(M/\msol)>10.48$
is consistent with a projected NFW profile.

\item We find that the number density of satellites depends  on the
stellar mass of the central galaxies. The most massive central
galaxies ($\log(M/\Msol)>10.78$) have $\sim$ 1.9 times the
number of satellites within 400 kpc compared to intermediate
mass centrals ($10.48 < \log(M_{\unit{c}}/\msol) < 10.78$), which is
significant at $\simeq$ 1.9$\sigma$.

\item For the most massive galaxies, $\log(M/\Msol)>10.78$, quiescent
centrals have $\sim$ 2 times the number of satellites within 400 kpc compared to star-forming centrals (significant at
$\simeq 3.1\sigma$). This excess persists at 2.7 sigma significance
even when we account for differences in the centrals' stellar mass
distributions.  In contrast, we find no significant difference in the
satellite distributions of less-massive quiescent and star-forming
centrals, $10.48<\log(M/\msol)<10.78$.  

\item We interpret the number density of satellites in our data using
the semi-analytic model of Guo et al. (2011) from the lightcone made
available by Henriques et al. (2012). We find that quiescent galaxies
in the model also have more satellites than star-forming galaxies of
similar stellar mass.  By matching the halo masses of the star-forming
and quiescent samples, we demonstrate that the difference in satellite
content in the simulation is due almost entirely to differences in
halo mass. We interpret this as evidence that the differences in
satellite content observed in the data is driven by a difference in
halo mass, and conclude that at stellar masses
$\log(M/\msol)>10.78$ the halos that host quiescent galaxies are
$\sim 0.3$ dex more massive than the halos that host star-forming
galaxies.

\item We use a simple model to investigate the relationship between
  quenching and halo mass, which roughly reproduces the observed
  quenched fractions and the differences in halo mass between
  star-forming and quenched galaxies in our two stellar mass bins. The
  model suggests a scenario where galaxies have some probability of
  being quenched over roughly two decades in halo mass,
  $\log(M_h/\msol)\sim 11-13.5$, where the probability increases with
  mass. This wide mass range suggests that, while halo mass quenching
  may be an important mechanism at $1 < z < 3$, halo mass is not the
  only factor driving quiescence. It remains unclear why some central
  galaxies in relatively massive halos can keep forming stars.

\end{itemize}

\acknowledgements

We wish to thank our collaborators in the \zfourge\ and CANDELS teams
for their dedication and assistance, without which this work would not
have been possible. We also wish to thank James Bullock and  Simon
White for valuable comments and feedback, and the anonymous referee
for a constructive report. Australian access to the  Magellan
Telescopes was supported through the National Collaborative  Research
Infrastructure Strategy of the Australian Federal Government. This work is supported by the National Science Foundation through grants AST-1009707 and AST-0808133. This work is based on observations taken by the CANDELS
Multi-Cycle Treasury Program with the NASA/ESA HST, which is operated
by the Association of Universities for Research in Astronomy, Inc.,
under NASA contract NAS5-26555.  This work is supported by HST program
number GO-12060.  Support for Program number GO-12060 was provided by
NASA through a grant from the Space Telescope Science Institute, which
is operated by the Association of Universities for Research in
Astronomy, Incorporated, under NASA contract NAS5-26555.   We
acknowledge generous support from the Texas A\&M University and the
George P.\ and Cynthia Woods Institute for Fundamental Physics and
Astronomy.

\bibliography{references}

\end{document}